\documentclass[%
reprint,
superscriptaddress,
 amsmath,amssymb,
 aps,
 citeautoscript,
prb,
floatfix,
]{revtex4-1}

\usepackage{graphicx}% Include figure files
\usepackage{dcolumn}% Align table columns on decimal point
\usepackage{bm}% bold math
\usepackage{comment}
\usepackage{color}
\usepackage{hyperref}% add hypertext capabilities
\usepackage[mathlines]{lineno}% Enable numbering of text and display math
\usepackage{gensymb}% Enable displaying degree of angle
\begin{document}

\title{Electronic stopping and proton dynamics in InP, GaP, and In$_{0.5}$Ga$_{0.5}$P from first principles}

\author{Cheng-Wei Lee}
\affiliation{Department of Materials Science and Engineering, University of Illinois at Urbana-Champaign, Urbana, IL 61801, USA}

\author{Andr\'e Schleife}
\email{schleife@illinois.edu}
\affiliation{Department of Materials Science and Engineering, University of Illinois at Urbana-Champaign, Urbana, IL 61801, USA}
\affiliation{Frederick Seitz Materials Research Laboratory, University of Illinois at Urbana-Champaign, Urbana, IL 61801, USA}
\affiliation{National Center for Supercomputing Applications, University of Illinois at Urbana-Champaign, Urbana, IL 61801, USA}

\date{\today}

\begin{abstract}
The phosphide-based III-V semiconductors InP, GaP, and In$_{0.5}$Ga$_{0.5}$P are promising materials for solar panels in outer space and radioisotope batteries, for which lifetime is a major issue.
In order to understand high radiation tolerance of these materials and improve it further, it is necessary to describe the early stages of radiation damage on fast time and short length scales.
In particular, the influence of atomic ordering, as observed e.g.\ in In$_{0.5}$Ga$_{0.5}$P, on electronic stopping is unknown.
We use real-time time-dependent density functional theory and the adiabatic local density approximation to simulate electronic stopping of protons in InP, GaP, and the CuAu-I ordered phase of In$_{0.5}$Ga$_{0.5}$P across a large kinetic energy range.
These results are compared to SRIM and we investigate the dependence on the channel of the projectile through the target.
We show that stopping can be enhanced or reduced in In$_{0.5}$Ga$_{0.5}$P and explain this using the electron-density distribution.
By comparing Ehrenfest and Born-Oppenheimer molecular dynamics, we illustrate the intricate dynamics of a proton on a channeling trajectory.
\end{abstract}

\maketitle

\section{\label{sec:Intro}Introduction}
Indium phosphide (InP) and In$_{0.5}$Ga$_{0.5}$P are well-suited materials for optoelectronic devices due to their direct (low-temperature) band gaps of 1.42 eV \cite{Pavesi:1991} and 1.99 eV \cite{Bugajski1983}, respectively.
Gallium phosphide (GaP) has an indirect gap of 2.34 eV at low temperature \cite{Panish1969}.
Using In$_{0.5}$Ga$_{0.5}$P, a tandem solar cell was demonstrated with an efficiency greater than 30\,\% in a double-junction \cite{Takamoto1997} and over 40\,\% in a triple-junction \cite{King2007} configuration.
In addition, In$_{0.5}$Ga$_{0.5}$P shows good resistance to energetic, charged-particle radiation, making it suitable for applications in extreme operational environments where lifetime is one of the major issues.
Examples include solar panels in outer space \cite{YAMAGUCHI2001,Dharmarasu2001} and radioisotope batteries \cite{Cress2006}.

Research devoted to analyzing degradation of solar panels caused by charged-particle radiation, typically relies on semi-classical models \cite{Yamaguchi1984,Yamaguchi1997,YAMAGUCHI2001,Dharmarasu2001} derived from the Shockley-Read \cite{Shockley1952} and Hall\cite{Hall1952} equation to describe recombination of electrons and holes in semiconductor devices.
This allowed attributing a gradual drop in efficiency of solar panels as fluence of radiation increases to decreased minority-carrier life times.\cite{Yamaguchi1984,Yamaguchi1995}
In addition, radiation-induced defects in InP based solar devices were found to be annealed by injection of minority carriers\cite{Dharmarasu2001,YAMAGUCHI2001} and the performance was partially recovered.
The enhanced annealing was attributed to the Bourgoin mechanism \cite{Yamaguchi1997,BOURGOIN1972}, i.e., a change of the charge state of defects due to injection that leads to faster diffusion. 
These insights illustrate that the semi-classical approach is useful for optimizing the design of devices, however, it has no access to atomic-scale details of the interaction between the charged projectile ions and the target material.
Such details are essential for understanding the underlying atomistic mechanisms.
Achieving this goal requires modern first-principles simulations such as the ones described here.

Previous studies \cite{ITOH1998,Bai2010,Klatt1993} showed that the defect dynamics in target materials exposed to charged-particle radiation differ between regions of bulk and interfaces, since interfaces can act as sink or source of defects.
Itoh reviewed the effect of interfaces on defect dynamics under the scenario of projectile kinetic energies that are too low to induce knock-on events \cite{ITOH1998}.
It was speculated that in this scenario, enhanced damage near interfaces can be attributed to stronger localization of excitons or slower recombination rates for Frenkel pairs \cite{ITOH1998}.
Furthermore, a recent study based on \textit{ab-initio} molecular dynamics \cite{Jiang2018} for primary knock-on events under particle radiation shows that cations in the GaAs/AlAs superlattice are more likely to be displaced than cations in pure GaAs or AlAs.
Therefore, it is critical to model the effect of interfaces on radiation damage.

Existing first-principles studies that aim at unraveling the effect of interfaces are limited to the linear-response approximation and focus on optical properties\cite{Botti2002,Botti2004} instead of electronic response to radiation.
Gumbs proposed an analytic expression for electronic stopping of a charged particle moving parallel to the surface of layered 2D free-electron gases, based on the random-phase approximation \cite{Godfrey1988}.
However, this approach is limited by the linear-response approximation and the specific geometric setup used in the derivation.
In particular, the charged projectile moves outside of the heterostructure and parallel to the surface.
Recently, Cruz combined the Bethe stopping theory\cite{Bethe1930} with a model of quantum confinement that imposes boundary conditions on the system, to study the effect of interfaces on electronic stopping \cite{Salvador2012}.
Although this method is not limited to a specific geometric setup, it still suffers from the linear-response approximation and the assumption of a fully ionized projectile as well as quantum confinement.

For device applications, high-quality In$_{0.5}$Ga$_{0.5}$P is fabricated, using molecular-beam epitaxy or organo-metallic vapor-phase deposition.
This leads to well-defined, atomically ordered phases \cite{Stringfellow1991}, instead of random solid solutions, with the ``CuAu-I'' ordered phase\cite{Kuan1985} being one simple example.
These ordered phases have different electronic and phonon band structures compared to solid solutions and to bulk materials, giving rise to different optical, electronic, and thermal properties \cite{Suzuki1988,Wei1994,Hassine1996,Ozolins1998,Duda2011,Chernyak1997}.
As discussed above, there are a few studies exploring materials response to particle radiation for interfaces in heterostructures where the components are much thicker than monolayers that are observed in atomically ordered phases.
However, due to the different geometry, these existing approaches cannot be applied to ordered phases irradiated by fast ions.
To the best of our knowledge, there is no literature on how ordered phases with periodicities on the single-monolayer scale affect the ultrafast electronic response to particle radiation.
This is the focus of the present work.

Here we use real-time time-dependent density functional theory (RT-TDDFT) to study the electronic response of InP, GaP, and In$_{0.5}$Ga$_{0.5}$P to highly energetic protons.
We compute the electronic stopping power and dynamics of the proton projectile for the individual materials.
Our results indicate that interfaces in In$_{0.5}$Ga$_{0.5}$P give rise to both local enhancement as well as reduction of instantaneous stopping, compared to pure InP or GaP.
We attribute this behavior to the redistribution of electron density caused by the formation of the ordered phase.
In addition, we compare the dynamics of the proton projectile using Ehrenfest and Born-Oppenheimer molecular dynamics.
Their difference suggests the importance of including non-adiabatic and excited-electron effects.

In Sec.\ \ref{sec:method} we summarize our computational approaches for ground-state calculations, real-time electron dynamics, and both average as well as instantaneous electronic stopping power.
In Sec.\ \ref{sec:ESP} and \ref{sec:IESP}, we report our results for average and instantaneous electronic stopping, respectively, for proton-irradiated InP, GaP, and In$_{0.5}$Ga$_{0.5}$P. 
In Sec.\ \ref{sec:dynamics}, we report the dynamics of a proton moving on a [100] channel using both Ehrenfest and Born-Oppenheimer molecular dynamics.
We compare the difference and discuss the importance to explicitly model electron dynamics. 
Lastly, we conclude and summarize our work in Sec.\ \ref{sec:conclusion}.

\section{\label{sec:method}Computational approach}

\subsection{\label{sec:gs}Ground-state calculations}

\begin{figure}
\includegraphics[width=0.8\columnwidth]{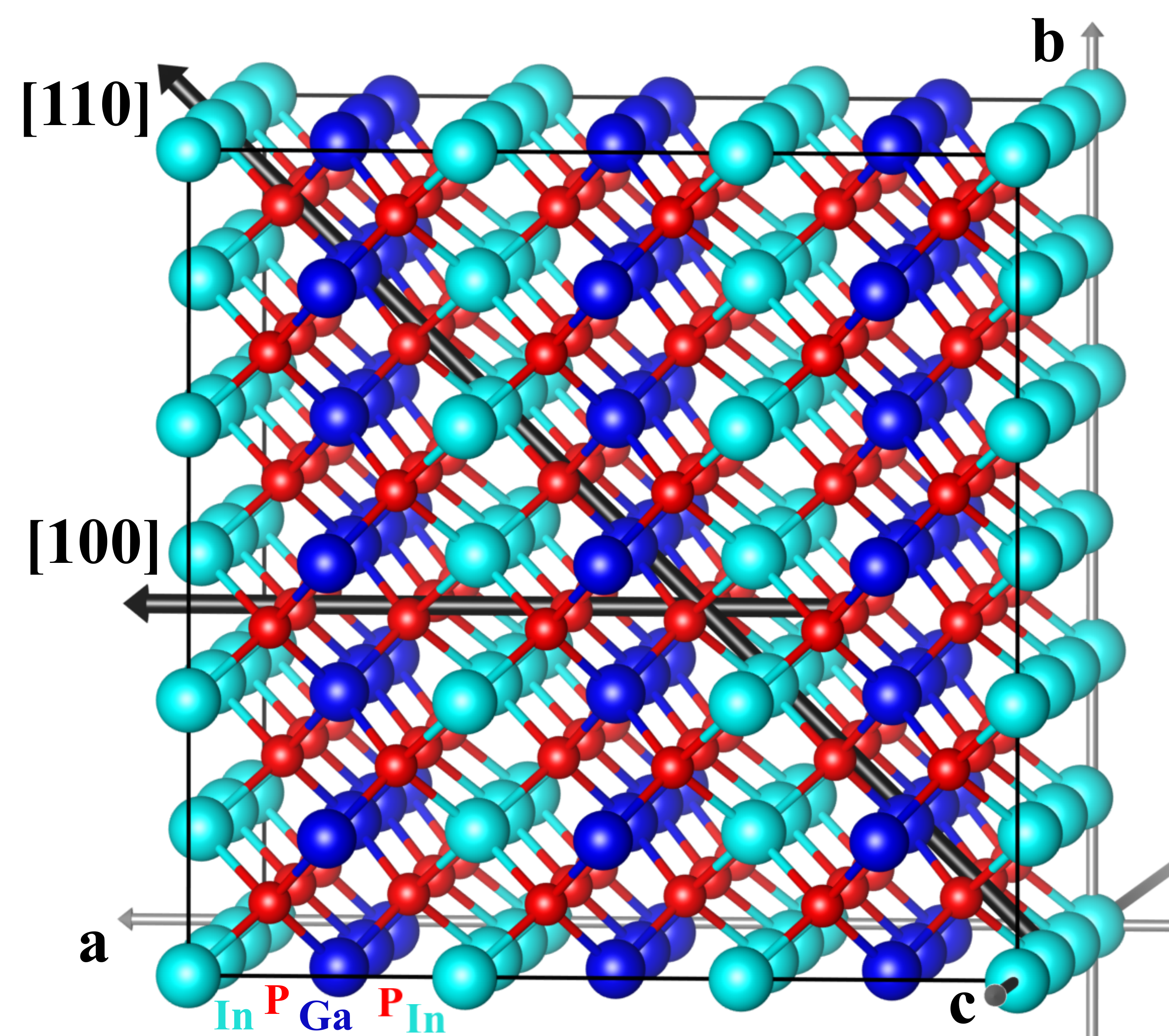}
\caption{\label{fig:superlattice}The 216-atom supercell used to represent In$_{0.5}$Ga$_{0.5}$P.
$a$, $b$, and $c$ are three orthogonal lattice axes.
Indium, gallium, and phosphorus are colored in light blue, dark blue, and red, respectively. 
Single layers of InP and GaP alternate along the [100] direction.
The two channeling [110] and [100] trajectories are shown as black arrows.
} 
\end{figure}

Using the Qb@ll code \cite{qbox_davis,qball2017}, we performed ground-state density functional theory (DFT) \cite{Hohenberg1964,Kohn1965} calculations for zinc-blende ($zb$) InP, $zb$-GaP, and the $zb$-based ordered CuAu-I phase \cite{Kim1999} of In$_{0.5}$Ga$_{0.5}$P (see Fig.\ \ref{fig:superlattice}).
On GaAs(001) substrates, the ``CuPt'' type atomic ordering of InGaP is more commonly observed \cite{Ueda1987,Bellon1988}, but the CuAu-I ordering was reported on GaAs(110) substrates before \cite{Kim1999}.
Hence, even though the CuAu-I phase is not the most common atomic ordering of In$_{0.5}$Ga$_{0.5}$P, it is chosen here as a reasonable and computationally feasible test case.
Kohn-Sham (KS) wave functions are expanded into a plane-wave basis with cutoff energies of 50 hartree ($E_{\mathrm{H}}$), 75 $E_{\mathrm{H}}$, and 75 $E_{\mathrm{H}}$ for InP, GaP, and In$_{0.5}$Ga$_{0.5}$P, respectively, to obtain total energies converged to within 0.184 m$E_{\mathrm{H}}$/atom.
The local-density approximation (LDA) is used to describe exchange and correlation \cite{Ceperley:1980,Perdew:1981} and the electron-ion interaction is described by norm-conserving Hamann, Schl{\"u}ter, and Chiang pseudopotentials as modified by Vanderbilt \cite{Vanderbilt:1985}.
We use pseudopotentials with 4$s^{2}$3$d^{10}$4$p^{1}$, 5$s^{2}$4$d^{10}$5$p^{1}$, and
3$s^{2}$3$p^{3}$ valence electrons for Ga, In, and P respectively.
The Brillouin zone is sampled using only the $\Gamma$ point, which is justified for the 216-atom supercells  used here.

Relaxed atomic geometries are computed using fits to the Murnaghan equation of state \cite{Murnaghan1944}.
This yields lattice constants of 11.07 and 10.24 $a_\mathrm{B}$ for InP and GaP respectively.
For In$_{0.5}$Ga$_{0.5}$P, we first determine the $a$/$c$ ration that gives similar pressure on all faces of the cell, and then scale the cell volume until the external pressure is below 0.5 GPa.
This yields cell dimensions $a$, $b$, and $c$ of 10.71 $a_\mathrm{B}$, 10.65 $a_\mathrm{B}$, and 10.65 $a_\mathrm{B}$, respectively.
All atomic positions are relaxed until forces are below 0.1 m$E_\mathrm{H}$/$a_\mathrm{B}$.

In order to isolate the effect of electronic excitations on ion dynamics, we also performed Born-Oppenheimer molecular dynamics (BOMD) simulations \cite{Marx:2009}.
Since the protons that represent particle radiation move very fast, smaller time steps compared to typical BOMD simulations were chosen.
This guarantees enough sampling points (210 points along the [100] trajectory) and conservation of energy. 
More specifically, a time step of 0.3 atomic units (at.\ u.) of time, 0.1 at.\ u., and 0.0375 at.\ u.\ is used for proton velocities of 0.5 at.\ u., 1.5 at.\ u., and 4.0 at.\ u., respectively.

\subsection{\label{sec:TDDFT}Real-time electron-ion dynamics}

We study real-time electron-ion dynamics using the Ehrenfest molecular dynamics approach \cite{Ehrenfest:1927,Marx:2009}.
Such simulations have become increasingly feasible even for solids \cite{Schleife:2014,Draeger:2017}, both due to the commendable balance of accuracy and computational efficiency of TDDFT \cite{Runge1984}, and due to the advent of modern supercomputers.
The electronic system is described by propagating time-dependent KS equations in real time using a fourth-order Runge-Kutta integrator \cite{Schleife:2012_c}.
A time step of 0.0145 at.\ u.\ was used and we verified that the electronic stopping power extracted from these simulations changed by less than 0.02 \% when the time step is halved.

Non-adiabatic electron-ion coupling is described by computing Hellman-Feynman forces from the time-dependent electron density \cite{Ehrenfest:1927,Marx:2009}.
These simulations are carried out using the TDDFT implementation within the Qb@ll code \cite{qbox_davis,qball2017,Schleife:2014,Draeger:2017}.

\subsection{\label{sec:stopping}Electronic stopping power}

When charged particles travel through a target compound, they transfer kinetic energy to that material \cite{Ziegler:1980}.
The energy loss ($dE$) per penetration depth ($dx$) is known as stopping power $S$ and has the unit of a force,
\begin{equation}
\label{eq:stopping}
S(x) = dE(x)/dx.
\end{equation}
As indicated in Eq.\ \eqref{eq:stopping}, stopping power is the instantaneous rate of energy transfer, e.g.\ from protons to the III-P compounds studied here.
In the low-kinetic energy regime, the projectile predominantly transfers energy to the ions of the target material (``nuclear stopping'').
However, for protons with kinetic energies higher than about 1 keV, more than 10 times as much energy is transferred from proton kinetic energy to the \emph{electronic} system of the III-P target material than to the ions (``electronic stopping'').
This electronic-stopping regime is the focus of this work. 

In Fig.\ \ref{fig:stopping} we compare electronic stopping for channeling, i.e.\ protons that travel on trajectories centered at [100] and [110] lattice channels, to off-channeling stopping geometries.
Our studies of off-channeling trajectories are motivated by experiment and enable us to compare to either amorphous or polycrystalline samples commonly used in practice.
Furthermore, even when the sample is a single crystal, experiment oftentimes studies off-channeling trajectories because standard Monte Carlo packages, such as SRIM \cite{Ziegler2010}, fail to predict damage and distribution of defects in target materials under channeling conditions \cite{Smith1991}.

In this work we follow the approach of simulating a random trajectory through the crystal, as devised in Ref.\ \onlinecite{Schleife2015}, to represent off-channeling protons.
For each velocity (projectile kinetic energy) we use a standard pseudo-random number generator to generate a random direction through the lattice.
In order to obtain results that are independent of the specific random direction, we ensure they are dissimilar from any lattice channel and each trajectory is simulated long enough to obtain convergence (see below).
We then fix the velocities of all atoms in the simulation, including the projectile, to exclude primary knock-on events \cite{Schleife2015}.
This also avoids numerical issues caused by very short distances between projectile and target atoms, for which large values of the Coulomb interaction would require much shorter time steps.
While this constitutes an approximation, it can be justified since the cross section for scattering between projectile and lattice atoms is very small for fast, light projectiles.
As discussed in detail in Ref.\ \onlinecite{Schleife2015}, this assumption of a frozen lattice is valid for high proton velocities such as the ones studied in this work, for which the time scale of interaction with the lattice is short.
This allows us to use the total-energy increase to compute electronic stopping for off-channeling protons \cite{Schleife2015}.
Full Ehrenfest dynamics simulations, where all ions are allowed to move according to Hellman-Feynman forces, are performed for channeling trajectories.

We compute averages of instantaneous electronic stopping for channeling projectiles by integrating over 2 lattice periods (unshaded area in Figs.\ \ref{fig:011_SP} and \ref{fig:001_SP}) after discarding the first half lattice period of a simulation, to avoid onset effects. 
Along the [100] and [110] directions, the 216-atom cell has three lattice periods but the length of the lattice period in [110] direction is by a factor of $\sqrt{2}$ larger than that in the [100] direction. 
Onset effects are obvious, e.g.\ in Fig.\ \ref{fig:001_SP}(a), where stopping near the onset is much larger than at later stages of the simulation.
Discarding also the last half lattice period of the simulation, allows us to mitigate the impact of excited electrons that re-enter the simulation cell due to periodic boundary conditions\cite{Schleife2015}.

As discussed in Ref.\ \onlinecite{Schleife2015}, the average electronic stopping for off-channeling projectiles is calculated from the instantaneous value using the slope of a linear regression fit to the $E(x)$ curve.
Initially, this result is sensitive to the trajectory length, however,
it eventually converges when the trajectory is long enough (approximately for a trajectory length of 200 $a_{\mathrm B}$). 

\section{\label{sec:results}Results and Discussion}

\subsection{\label{sec:ESP}Average electronic stopping}

\begin{figure}
\includegraphics[width=0.9\columnwidth]{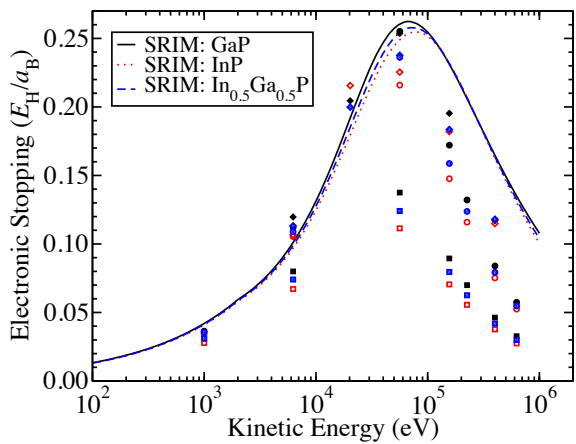}
\caption{\label{fig:stopping}Electronic stopping of InP (red open), GaP (black filled), and In$_{0.5}$Ga$_{0.5}$P (blue partial filled) under proton irradiation. [100] (circles), [110] (squares), and random trajectories (diamonds) are compared with results computed using ``The Stopping and Range of Ions in Matter'' (SRIM) \cite{Ziegler2010} (lines). 
}
\end{figure}

In Fig.\ \ref{fig:stopping} we show the dependence of the electronic stopping power on proton kinetic energy as computed from our first-principles simulations.
This figure compares two channeling proton trajectories to the off-channeling configuration for GaP, InP, and In$_{0.5}$Ga$_{0.5}$P.
From this comparison, it becomes immediately clear that electronic stopping in all three III-P compounds depends strongly on the trajectory:
For all proton kinetic energies, the [110] channel leads to the smallest electronic stopping.
The [100] channel shows similar electronic stopping as the off-channeling trajectory before the stopping maximum, but also leads to smaller stopping close to and even more so after the peak of the curve.

The first observation of smaller stopping along the [110] channel can be explained by the effective electron density that the projectile interacts with along this trajectory.
When protons travel on a [110] channel, the average distance between the proton and first-nearest-lattice atoms is about 50\,\% longer than for protons on a [100] channel.
Since most of the electron density is located near the ions, protons on [110] channels interact with smaller electron density, leading to weaker electronic stopping.
This finding is consistent with a previous RT-TDDFT study of proton-irradiated germanium \cite{Ullah:2015} and a study based on scattering theory for energy loss in a non-uniform electron gas \cite{Winter:2003}.

The second observation, that off-channeling projectiles lead to higher stopping than channeling projectiles, has been reported in the literature before and was attributed to stopping contributions from semi-core electrons \cite{Schleife2015, Quashie:2016}.
In order for semi-core electrons to contribute to electronic stopping, protons need to have high enough kinetic energy to excite the semi-core states.
In addition, these excitations require spatial proximity of the proton and semi-core wave functions, i.e.\ very small distances between proton and ions, which we only capture by random trajectories in our simulations.

Finally, comparison of our results to data that we computed using ``The Stopping and Range of Ions in Matter'' (SRIM) \cite{Ziegler2010} shows good overall agreement and confirms our interpretation.
Since SRIM assumes an amorphous structure of the target material, the large range of electron density values that a projectile experiences as it traverses an amorphous target is most closely represented by our off-channeling trajectory.
Consequently, when comparing our results for off-channeling electronic stopping to SRIM, we find good agreement before the electronic-stopping peak, but deviations become significant especially for higher kinetic energies.
This behavior has been identified in the literature\cite{Schleife2015,Yost:2017} before and one possible explanation invokes electronic-stopping contributions due to semi-core electrons that are missing in the pseudopotentials used here (see supplemental material).
Another limitation is the use of the adiabatic LDA in this work, and, while this a topic of ongoing research \cite{Nazarov:2007}, it is currently unknown how this quantitatively affects electronic stopping of protons.
Finally, the simulation cell size also affects the accuracy of plasma excitations since it limits the maximum wave length for a plasmon in the simulation \cite{Correa:2018}.

\begin{table*}
\caption{\label{tab:stopping}Electronic stopping (in $E_\mathrm{H}/a_\mathrm{B}$) as a function of projectile velocity $v$ (in at.\ u.) for GaP, InP, and In$_{0.5}$Ga$_{0.5}$P and [100] channel/[110] channel/off-channeling.
Fewer off-channeling cases were studied due to the larger computational cost of obtaining converged results.
Since our results deviate from SRIM data near the maximum of electronic stopping, an additional velocity slightly below ($v$=0.9 at.\ u.) was chosen for off-channeling protons.
We also compare averages of electronic stopping for InP and GaP with In$_{0.5}$Ga$_{0.5}$P. $\Delta$ is the stopping power difference of In$_{0.5}$Ga$_{0.5}$P from the average value of InP and GaP, divided by that average. 
Relative errors are less than 5\,\%, when estimated from averages over different lattice periods for channeling projectiles (see supplemental material for details).
}
\begin{tabular}{|c||c|c|c||c|c|c||c|c|c||c|c|c||c|c|c|}
\hline
$\boldmath{v}$ & \multicolumn{3}{c||}{GaP} & \multicolumn{3}{c||}{InP} & \multicolumn{3}{c||}{Avg.} & \multicolumn{3}{c||}{In$_{0.5}$Ga$_{0.5}$P} & \multicolumn{3}{c|}{$\Delta$ (\%)} \\ \hline
0.2 & 0.0365 & 0.0326 & --- & 0.0345 & 0.0277 & ---  & 0.0355 & 0.0302 & --- & 0.0357 & 0.0309 & --- & 0.71 & 2.41 & --- \\ \hline
0.5 & 0.1121 & 0.0800 & 0.1197 & 0.1053 & 0.0671 & 0.1066 & 0.1087 & 0.0735 & 0.1131  & 0.1089 & 0.0740 & 0.1132 & 0.18 & 0.69 & 0.07 \\ \hline
0.9 & --- & --- & 0.2045 & --- & --- & 0.2156 & --- & --- & 0.2101 & --- & --- & 0.1999 & --- & --- & $-4.86$ \\ \hline
1.5 & 0.2552 & 0.1375 & 0.2537 & 0.2198 & 0.1114 & 0.2254 & 0.2375 & 0.1244 & 0.2395  & 0.2362& 0.1241 & 0.2377 & $-0.55$ & $-0.30$ & $-0.78$ \\ \hline
2.5 & 0.1721 & 0.0894 & 0.1954 & 0.1463 & 0.0705 & 0.1820 & 0.1588 & 0.0799 & 0.1887 & 0.1587 & 0.0795 & 0.1835 & $-0.03$ & $-0.53$ & $-2.73$ \\ \hline
3.0 & 0.1327 & 0.0700 & ---  & 0.1145 & 0.0556 & --- & 0.1226 & 0.0628 & --- & 0.1238 & 0.0625 & --- & $0.16$ & $-0.39$ & --- \\ \hline
4.0 & 0.0839 & 0.0462 & 0.1170 & 0.0741 & 0.0377 & 0.1149 & 0.0790 & 0.0420 & 0.1160 & 0.0793 & 0.0419 & 0.1181 & $0.38$ & $-0.25$ & 1.81 \\ \hline
5.0 & 0.0574 & 0.0327 & --- & 0.0518 & 0.0275 & --- & 0.0545 & 0.0301 & --- & 0.0548 & 0.0300 & --- & $0.55$ & $-0.34$ & --- \\ \hline
\end{tabular}
\end{table*}

We also note that our results agree with SRIM regarding the relative magnitude of electronic stopping across the different materials.
Except for off-channeling projectiles with $v$=0.9 at.\ u.\
we consistently find stopping in GaP to be the largest,  in InP to be the smallest, and in In$_{0.5}$Ga$_{0.5}$P to be in between.
More specifically, we find that electronic stopping of In$_{0.5}$Ga$_{0.5}$P is very close to the average of stopping in InP and GaP.
The data in Table \ref{tab:stopping} illustrates that the relative differences are below 1\,\% across most of the velocities for the [100] and [110] channels.
This also holds for the density of valence electrons (see Sec.\ \ref{sec:method}) for these compounds:
That of In$_{0.5}$Ga$_{0.5}$P is 4.00\,$\times$\,10$^{23}$ cm$^{-3}$, which is within 1.3\,\% of the average of 4.52\,$\times$\,10$^{23}$ cm$^{-3}$ (GaP) and 3.58\,$\times$\,10$^{23}$ cm$^{-3}$ (InP).
we assume to zeroth order that a proton moving on a channel through In$_{0.5}$Ga$_{0.5}$P interacts half of the time with InP-like electron density and half of the time with GaP-like electron density.
Within the Lindhard model, electronic stopping is proportional to the electron density \cite{Lindhard:1964}, hence, we conclude that this model and the linear approximation describe electronic stopping for channeling in the CuAu-I ordered phase of In$_{0.5}$Ga$_{0.5}$P very well.
We will refine this picture below, using the actual electron-density distribution in In$_{0.5}$Ga$_{0.5}$P.

As described above, for off-channeling projectiles we use different random trajectories for the different velocities and III-P compounds.
Due to the statistical nature of this approach, convergence is computationally challenging:
While each trajectory converges to good accuracy after about 200 $a_\mathrm{B}$, a given random trajectory may represent a good cell average only after much longer lengths.
We observe this for $v$=0.9 at.\ u., where Fig.\ \ref{fig:stopping} shows a different relative ordering for the different materials, compared to the other velocities.
The InP trajectory in this case more often samples close proximity to semi-core electrons and, thus, higher stopping (see supplemental material).
Much longer runs would be required to eliminate this influence from the final stopping result.

From previous electronic-structure calculations\cite{Wei1994,SEDDIKI2013} it is expected that formation of an ordered phase results in breaking of translational symmetry and, therefore, splitting and energy shifting of bands and states inside the band gap.
However, our results indicate, that after averaging over instantaneous stopping, the local electronic structure of In$_{0.5}$Ga$_{0.5}$P has a very minor influence on electronic stopping.
We attribute this to the large projectile velocities studied in this work, compared to the changes in the electronic structure.
The situation is different for instantaneous stopping, which we discuss next.

\subsection{\label{sec:IESP}Instantaneous electronic stopping}

\begin{figure}
\includegraphics[width=0.92\columnwidth]{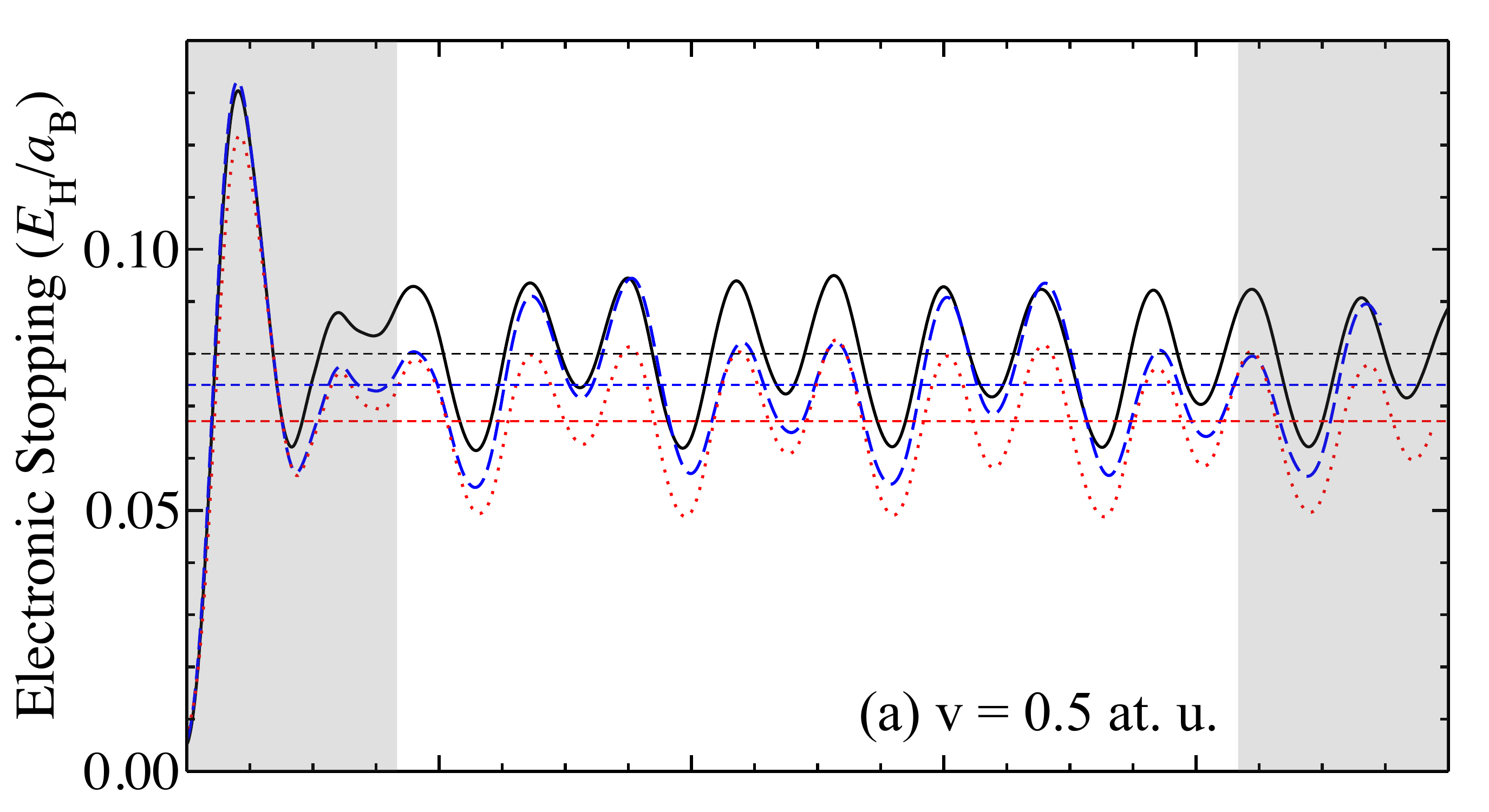}
\includegraphics[width=0.92\columnwidth]{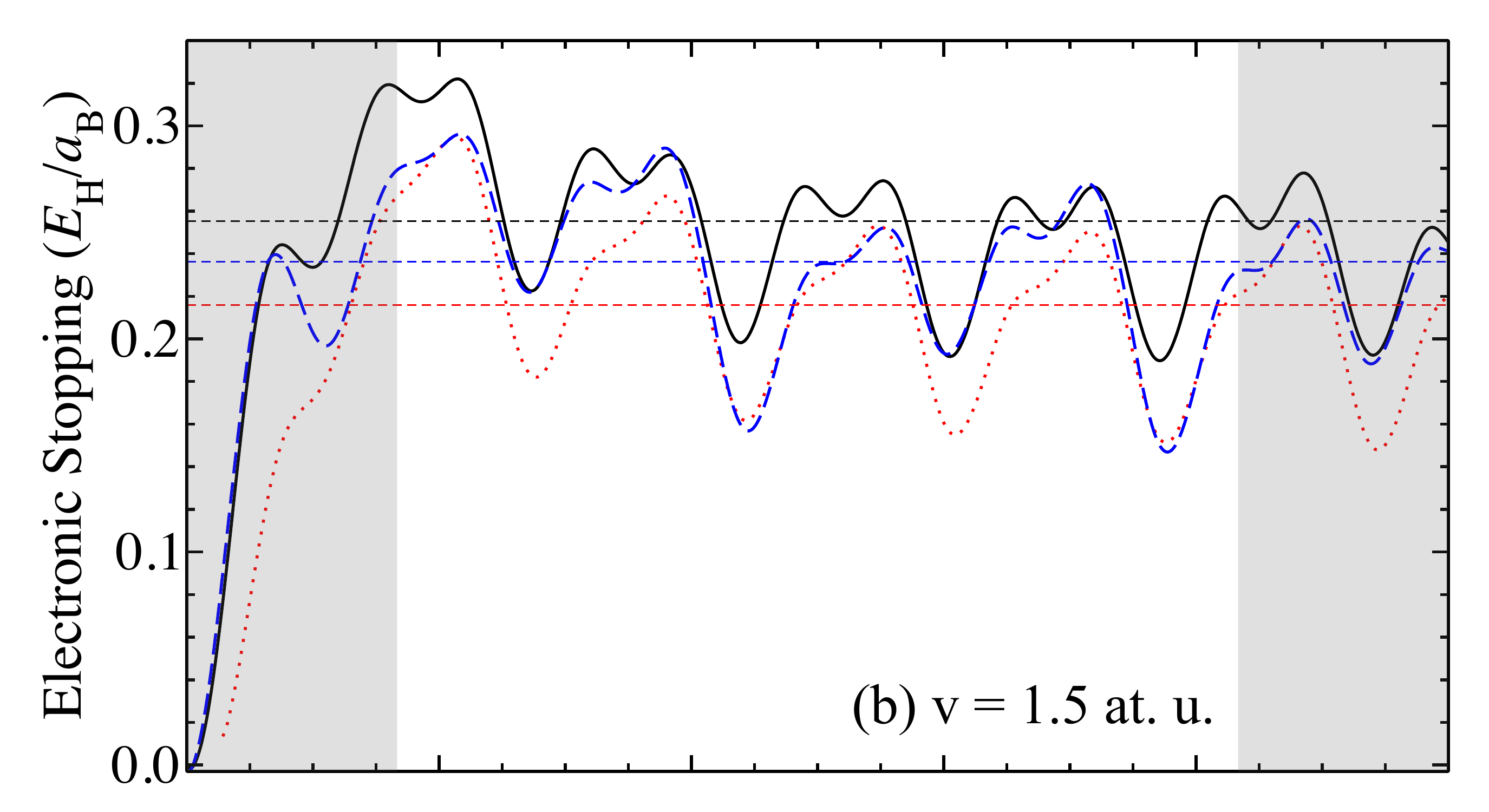}
\includegraphics[width=0.92\columnwidth]{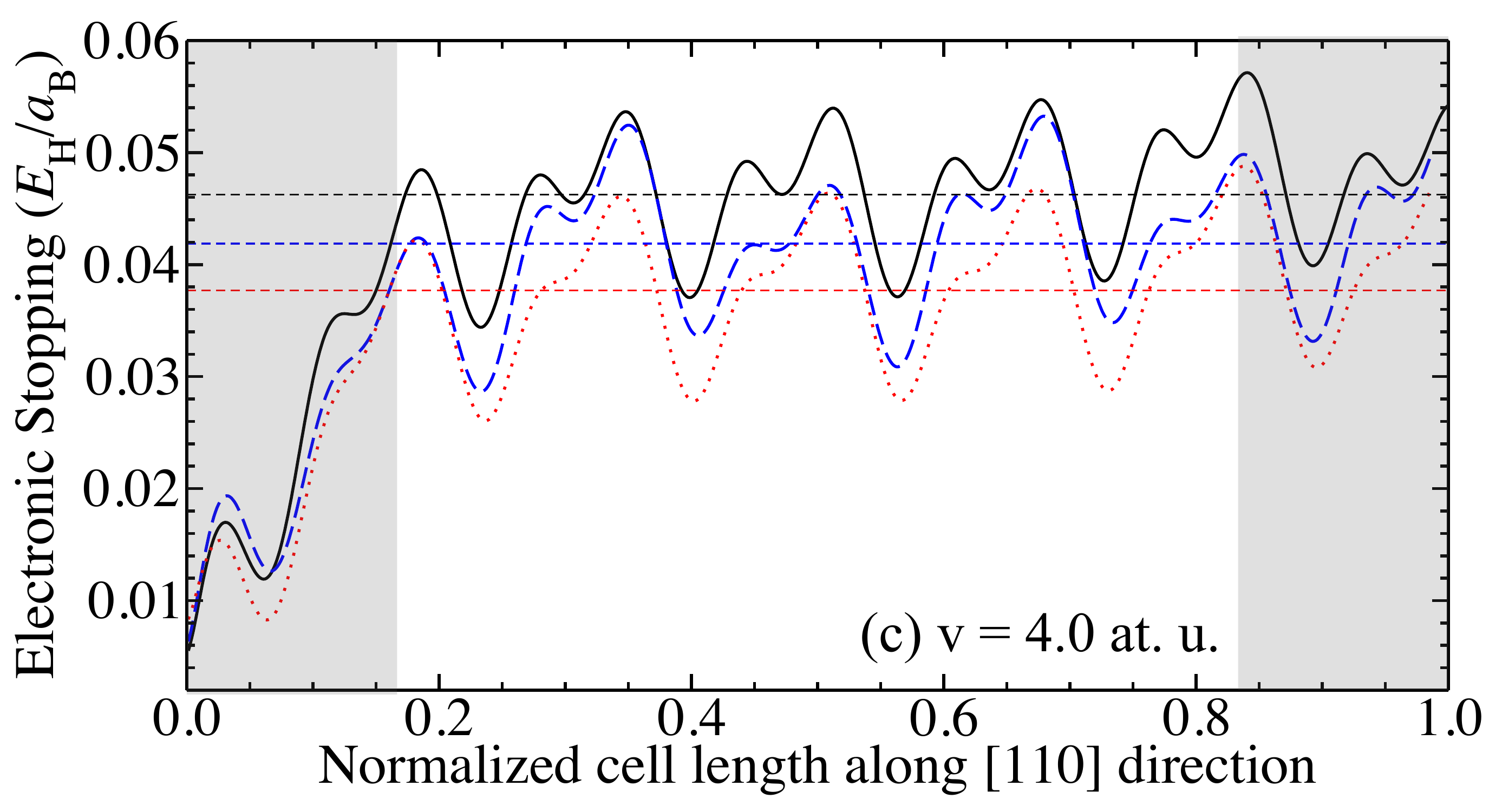}
\caption{\label{fig:011_SP}Instantaneous electronic stopping for a proton on a [110] channel with a velocity of (a) 0.5 at.\ u., (b) 1.5 at.\ u., and (c) 4.0 at.\ u.
Red dotted, black solid, and blue dashed lines are InP, GaP, and In$_{0.5}$Ga$_{0.5}$P, respectively. 
Horizontal dashed lines represent the corresponding average electronic stopping, computed for the unshaded part of the trajectory (see text).
}
\end{figure}

\begin{figure}
\includegraphics[width=0.92\columnwidth]{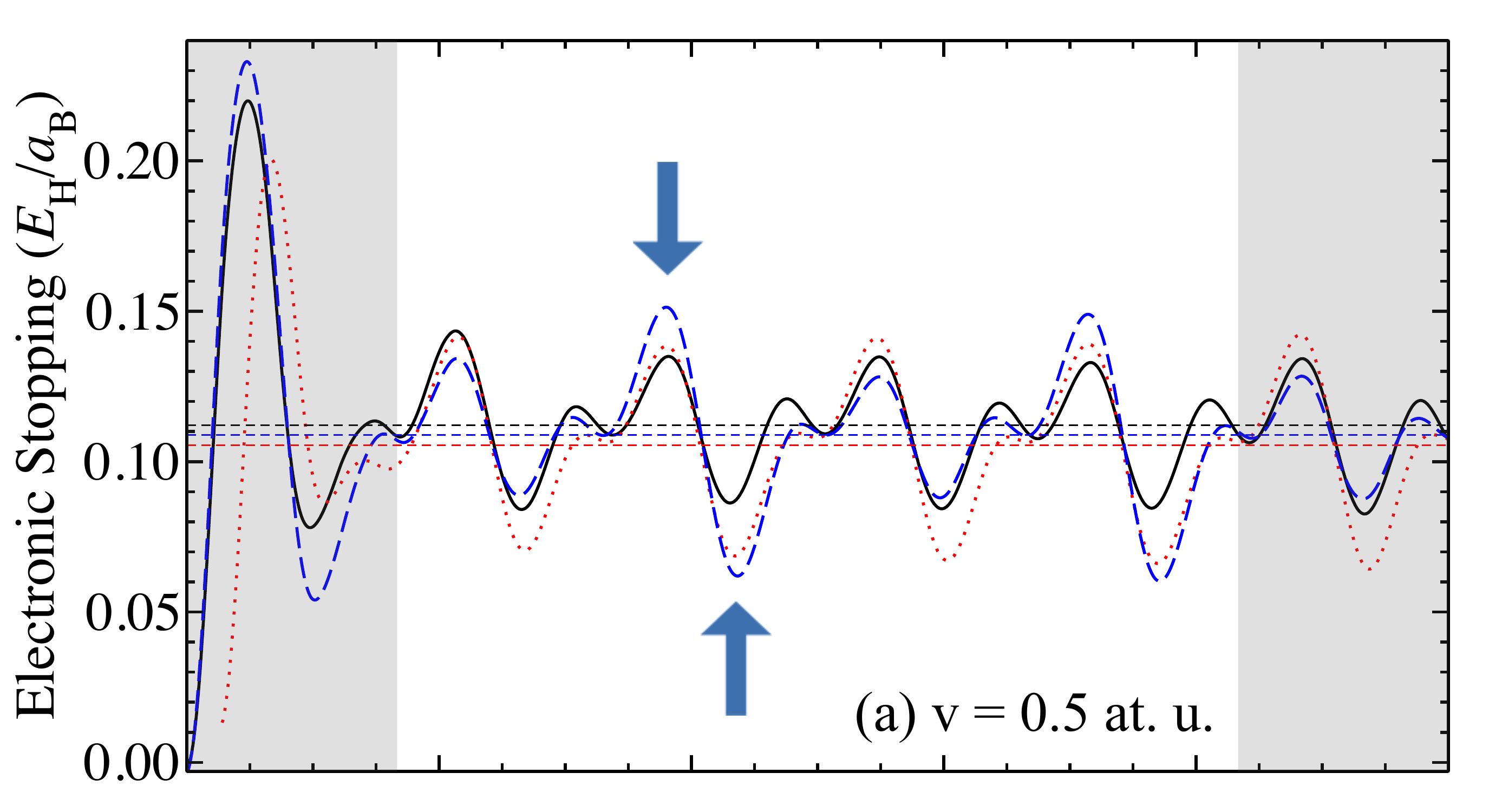}
\includegraphics[width=0.92\columnwidth]{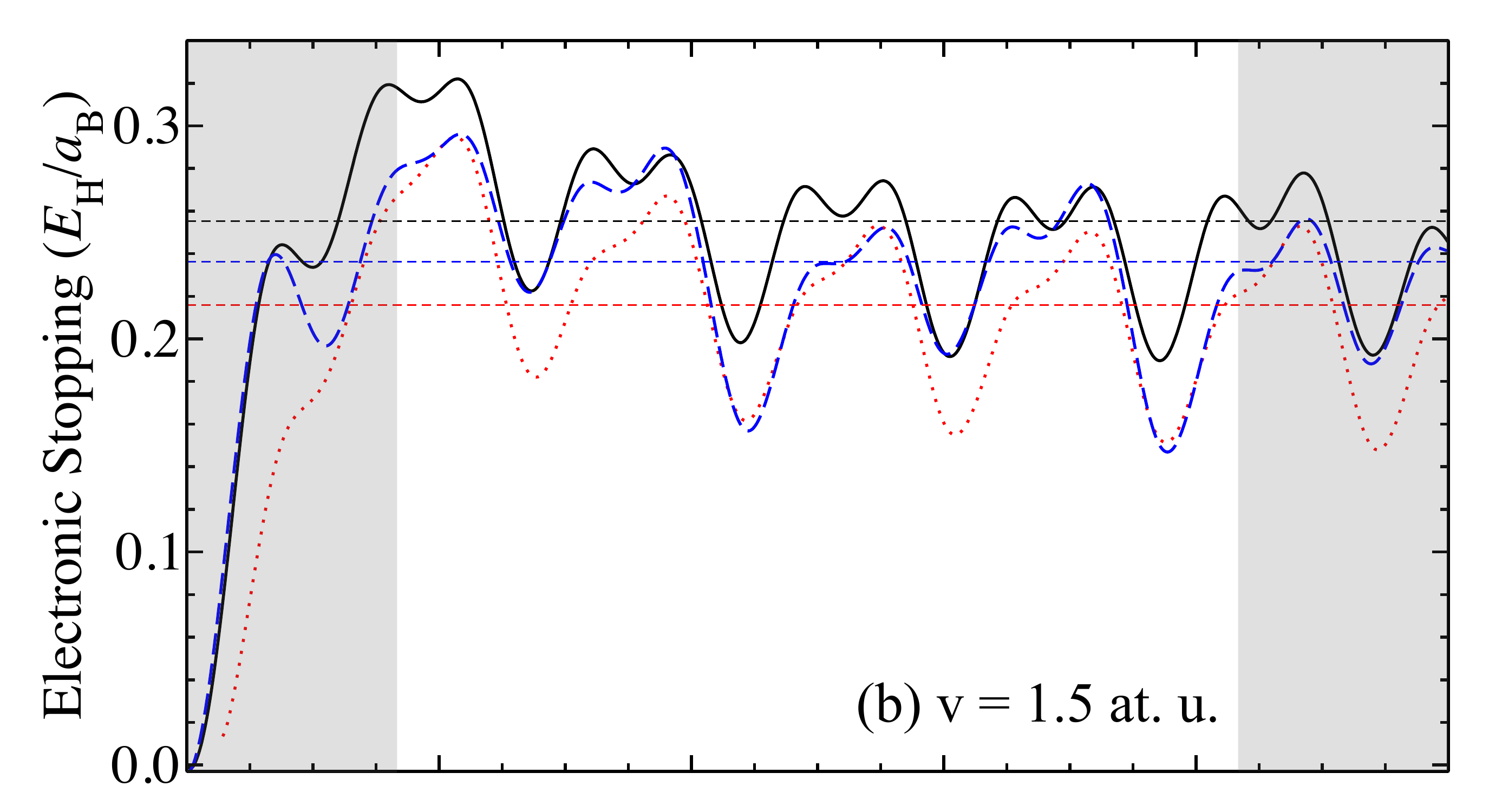}
\includegraphics[width=0.92\columnwidth]{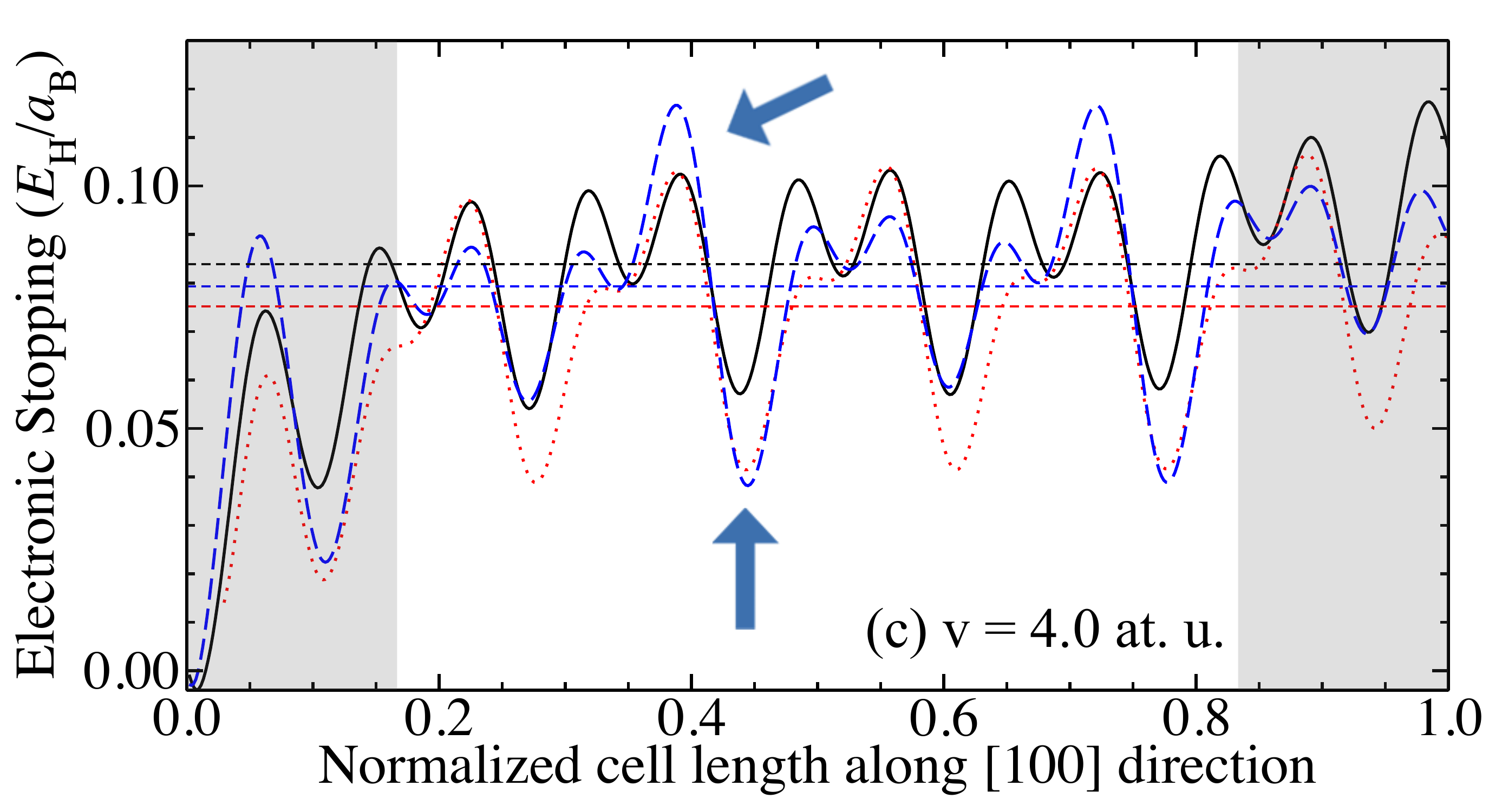}
\caption{\label{fig:001_SP}Instantaneous electronic stopping for a proton on a [100] channel with a velocity of (a) 0.5 at.\ u., (b) 1.5 at.\ u., and (c) 4.0 at.\ u.
Red dotted, black solid, and blue dashed lines are InP, GaP, and In$_{0.5}$Ga$_{0.5}$P, respectively. 
Horizontal dashed lines represent the corresponding average electronic stopping, computed for the unshaded part of the trajectory (see text).
Blue arrows indicate local enhancement/reduction.
}
\end{figure}

\begin{figure}
\includegraphics[width=0.32\textwidth]{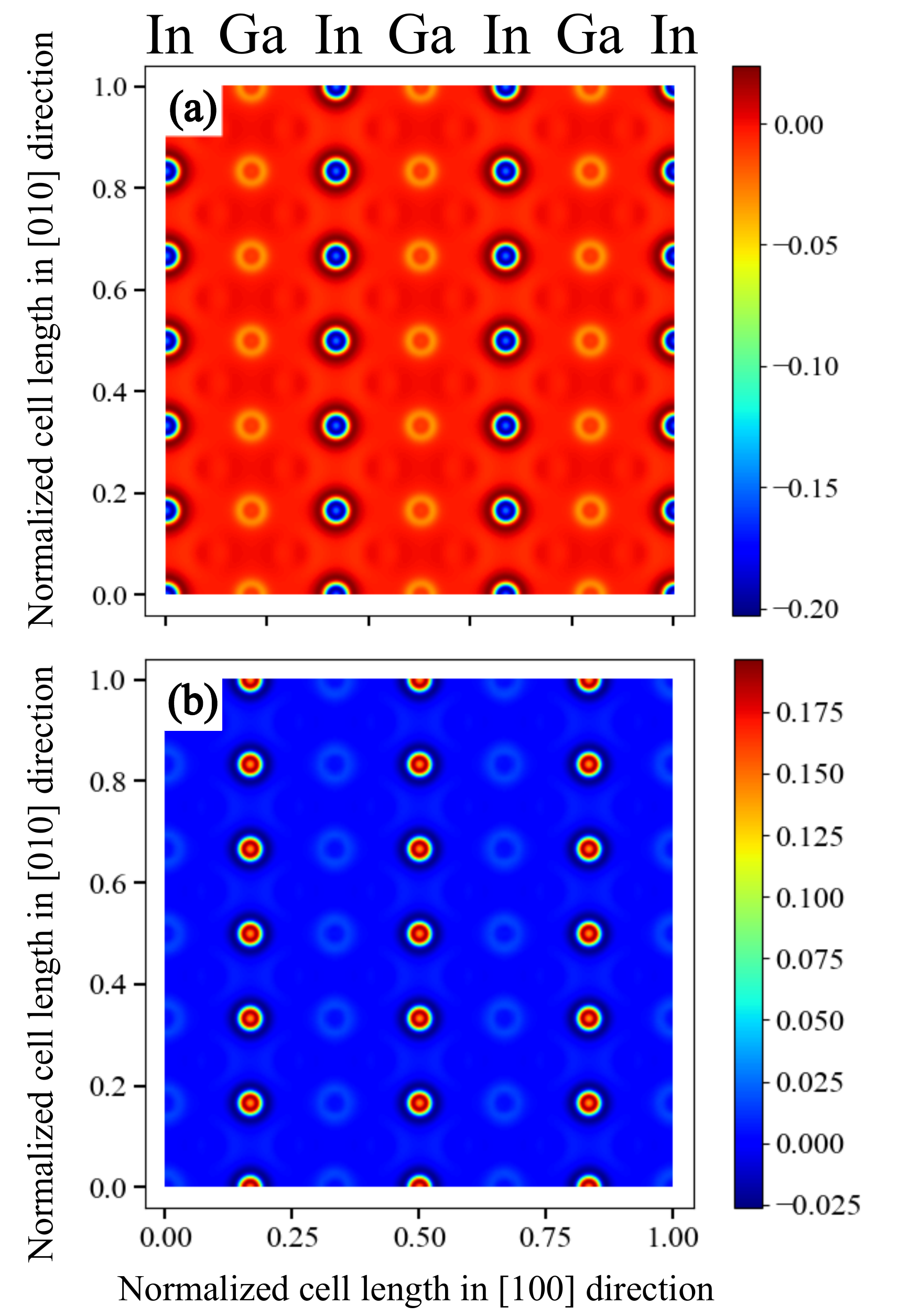}
\caption{\label{fig:rho_diff}Difference of the average of the electron density along [001] direction between In$_{0.5}$Ga$_{0.5}$P and (a) GaP or (b) InP.
Layers of In and Ga atoms are labeled.
In oder to compare the different III-P compounds, the cell length is normalized, putting cations and P atoms in the same relative positions.
(a) shows that in In$_{0.5}$Ga$_{0.5}$P there is less electron density near Ga ions than in GaP and (b) shows that there is more electron density around In atoms in In$_{0.5}$Ga$_{0.5}$P, compared to InP.
The difference in electron density near P atoms is small and, thus, hardly visible.
}
\end{figure}

Our RT-TDDFT results unambiguously show that instantaneous electronic stopping reveals a dependency on the local environment.
Since all the III-P compounds have slightly different cell parameters, we use the normalized cell length for InP, GaP, and In$_{0.5}$Ga$_{0.5}$P in order to help visualization and comparison (see Figs.\ \ref{fig:011_SP}, \ref{fig:001_SP}, and \ref{fig:rho_diff}).
This ensures that the same local environment is compared for all the III-P compounds.
Figure \ref{fig:011_SP} illustrates that instantaneous electronic stopping of protons moving with three different velocities on a [110] channel in In$_{0.5}$Ga$_{0.5}$P oscillates between InP-like and GaP-like behavior.
When the proton is near the InP layer of In$_{0.5}$Ga$_{0.5}$P, it locally follows the curve of InP and, similarly, when it is near the GaP layer it follows GaP stopping.
After averaging instantaneous stopping along the trajectory as discussed above, we then find that average stopping in In$_{0.5}$Ga$_{0.5}$P is very close to the average of GaP and InP electronic stopping.

For protons on a [100] channel, however, we find a totally different behavior and even a velocity dependence, as illustrated in Fig.\ \ref{fig:001_SP}.
As can be seen from this figure, for velocities of 0.5 at.\ u.\ and 4.0 at.\ u., the instantaneous stopping of In$_{0.5}$Ga$_{0.5}$P is locally larger or smaller than that of InP and GaP.
For these two velocities, the ordered phase of In$_{0.5}$Ga$_{0.5}$P gives rise to local enhancement and reduction of electronic stopping.
However, in the case of a proton with a velocity of 1.5 at.\ u.\ the stopping is again within the boundaries defined by InP and GaP, similar to what we discussed above for the [110] channel.
We attribute this velocity dependence to electronic states that appear in the ordered In$_{0.5}$Ga$_{0.5}$P phase and that lead to the observed behavior.

The ground-state electron density allows to analyze this in more detail and we find that its spatial distribution in In$_{0.5}$Ga$_{0.5}$P contributes to the local enhancement and reduction.
To illustrate this, Fig. \ref{fig:rho_diff} shows the difference of the electron-density average along the [001] direction between In$_{0.5}$Ga$_{0.5}$P and GaP as well as InP as a 2D plot.
The top panel shows that in In$_{0.5}$Ga$_{0.5}$P there is less charge around Ga ions than in GaP, and the bottom panel shows that there is more charge around In atoms in In$_{0.5}$Ga$_{0.5}$P, compared to InP.
The difference for P atoms is negligible.
Comparing this to the data in Fig.\ \ref{fig:001_SP} illustrates that enhanced stopping occurs near In atoms and reduced stopping is observed near Ga atoms for a proton on a [100] channel, which matches the behavior of the electron density near these atoms.
Contrary, the proton on a [110] channel is further away from these atoms and does not sample these electron-density differences.
Hence, no local enhancement or reduction of electronic stopping is observed in Fig.\ \ref{fig:011_SP}.
Our observation that not all proton velocities lead to enhancement or reduction of electronic stopping cannot be understood in this model.
Instead, we conjecture that this is related to the specific electronics states in In$_{0.5}$Ga$_{0.5}$P that are responsible also for the electron-density differences discussed above.

\subsection{\label{sec:dynamics}Dynamics of a proton on a [100] channel in In$_{0.5}$Ga$_{0.5}$P}

\begin{figure}
\includegraphics[width=0.92\columnwidth]{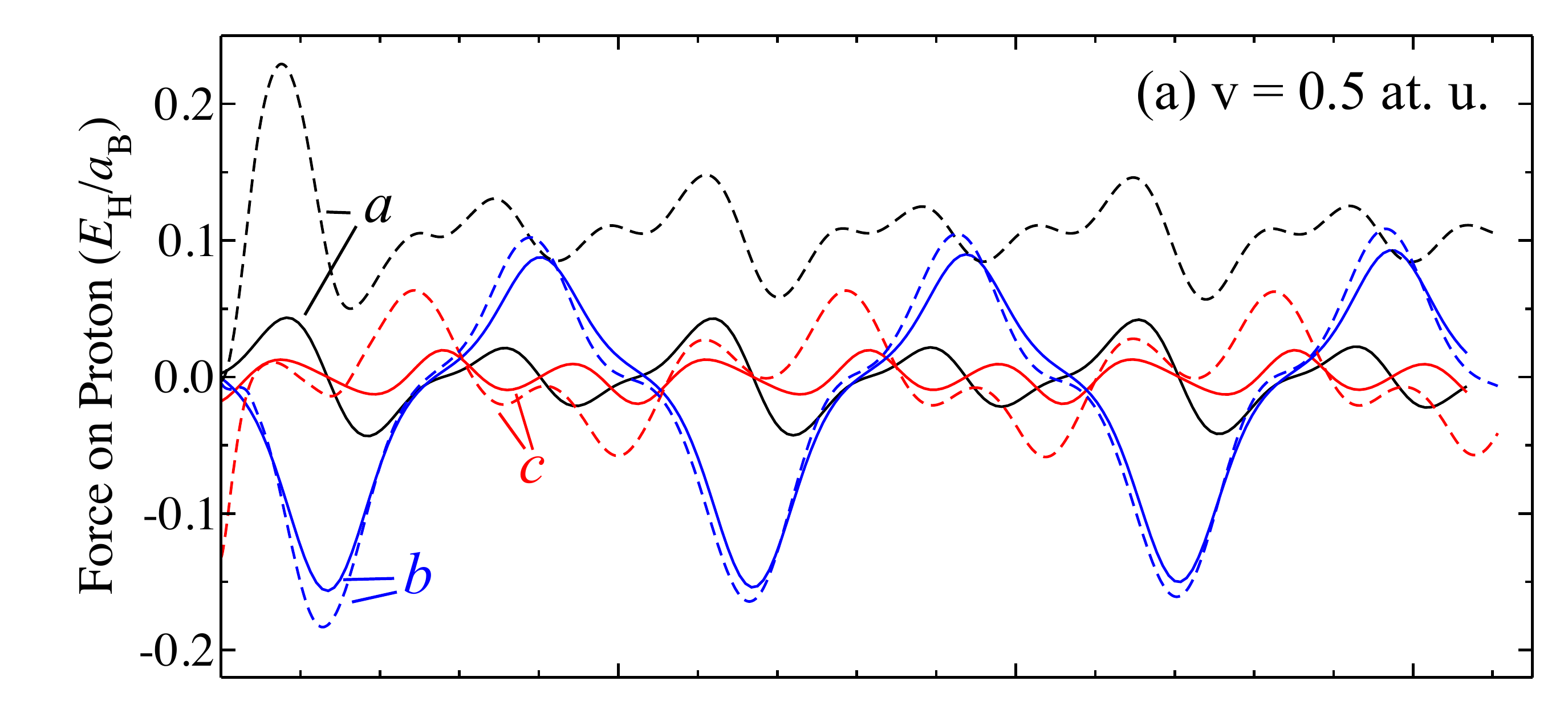}
\includegraphics[width=0.92\columnwidth]{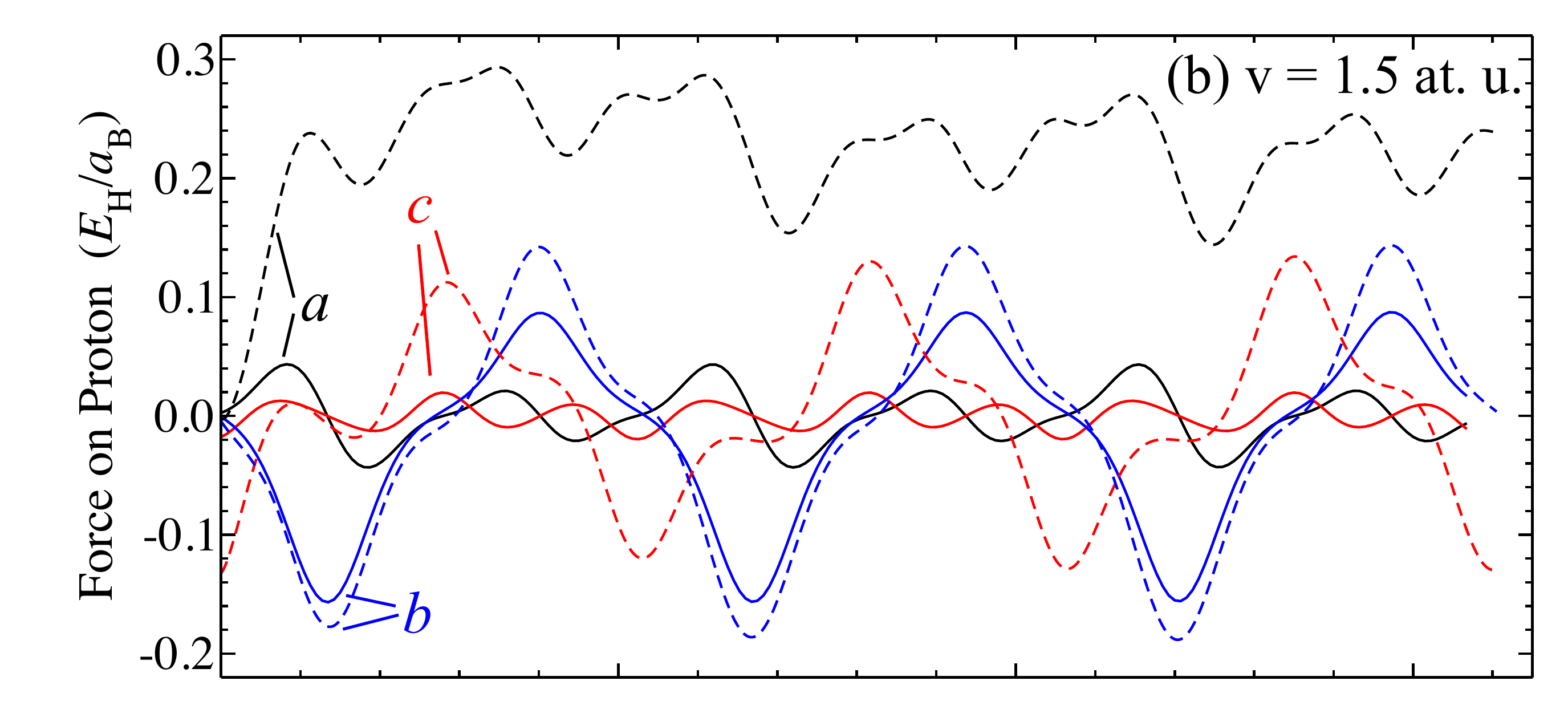}
\includegraphics[width=0.92\columnwidth]{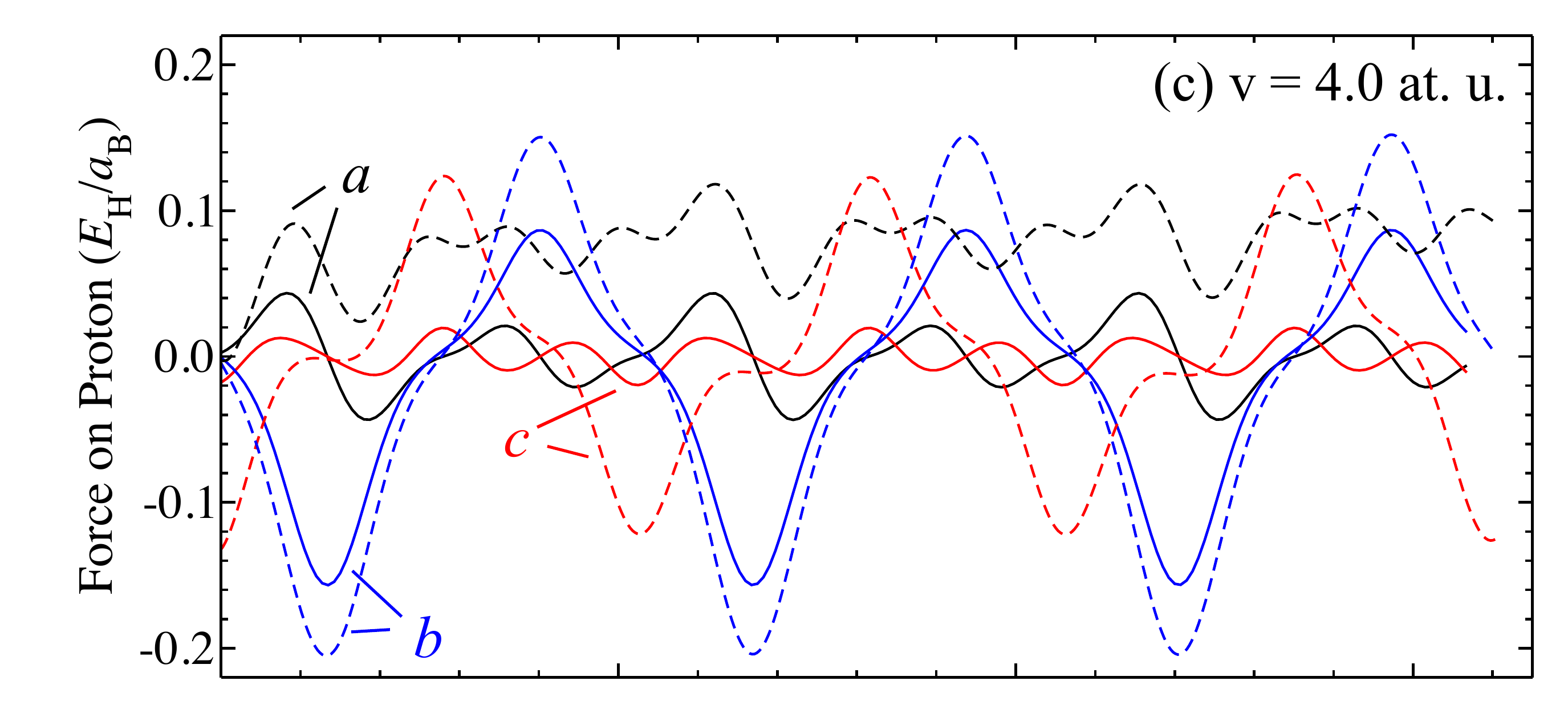}
\includegraphics[width=0.92\columnwidth]{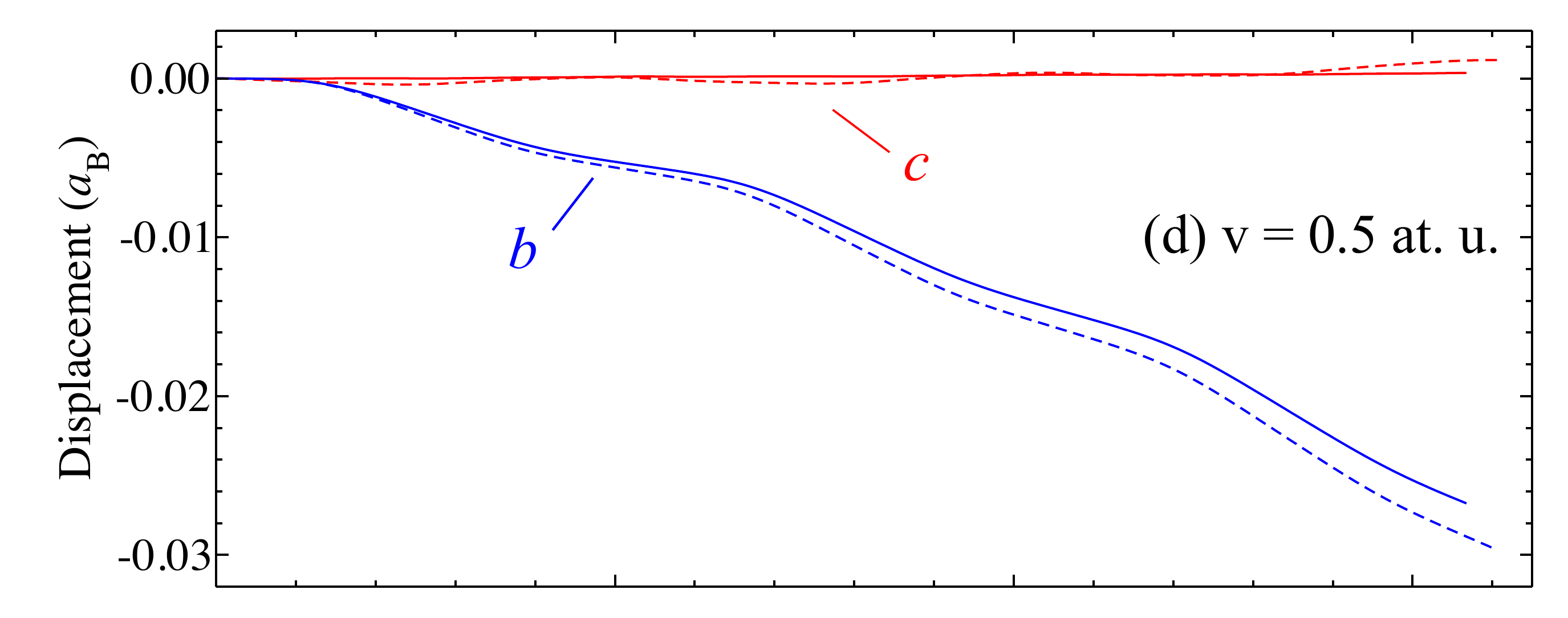}
\includegraphics[width=0.92\columnwidth]{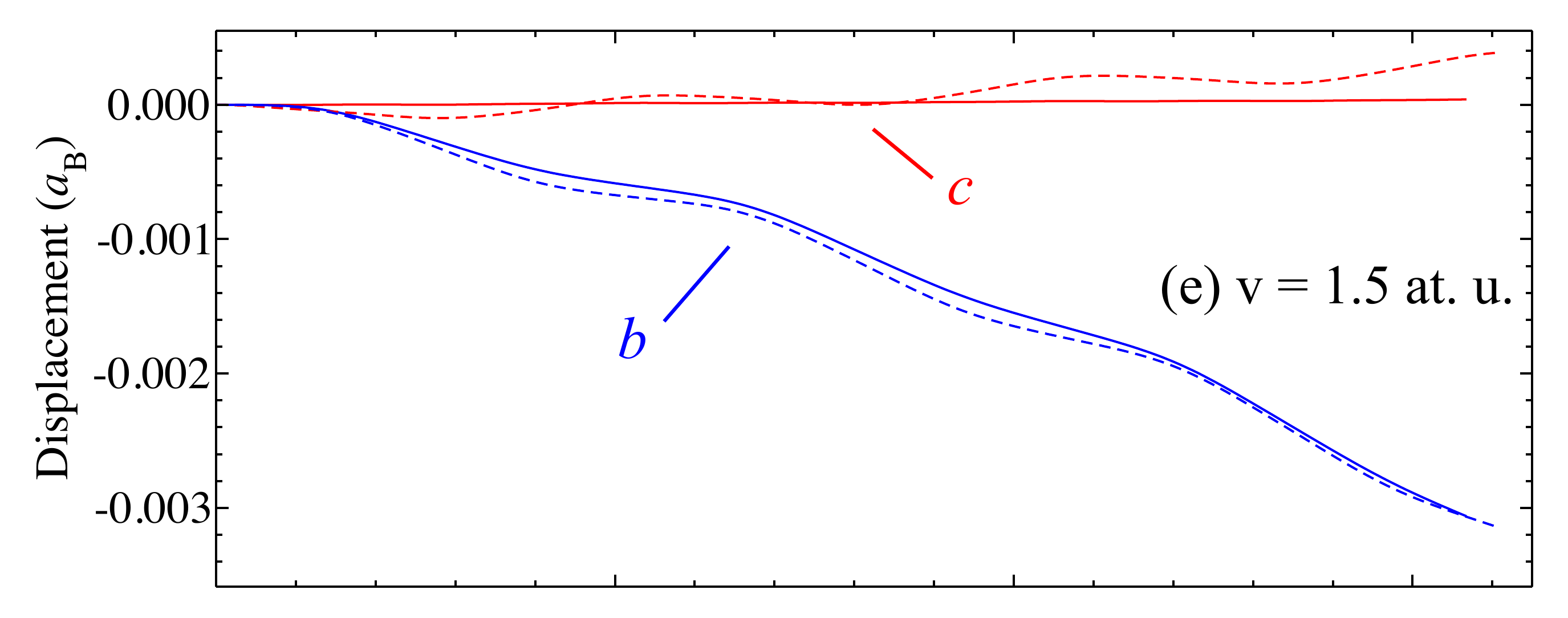}
\includegraphics[width=0.92\columnwidth]{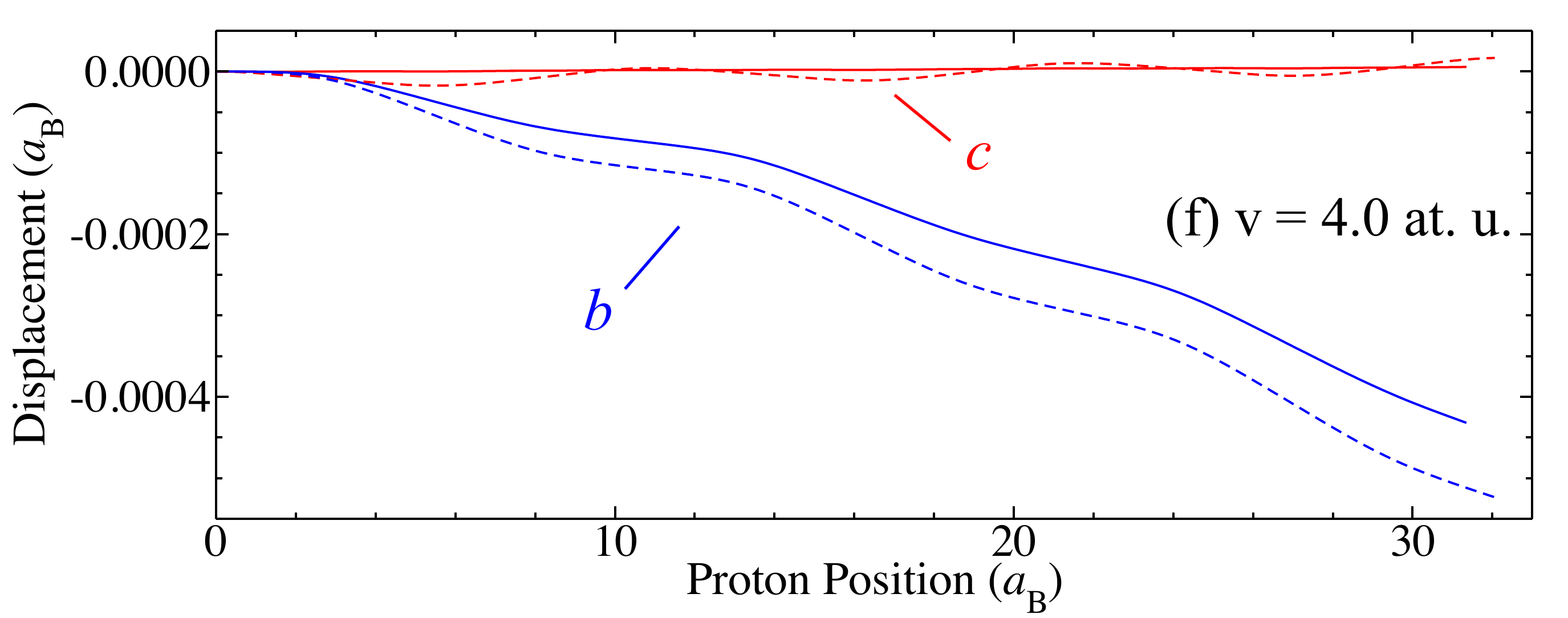}
\caption{\label{fig:Dyns_BOMD_vs_ED}
Dynamics of a proton on a [100] channel in In$_{0.5}$Ga$_{0.5}$P. 
Black, blue, and red correspond to the $a$, $b$, and $c$ components of force and displacement, as defined in Fig.\ \ref{fig:superlattice}. Solid and dashed lines correspond to BOMD and Ehrenfest dynamics, respectively.
Forces and displacements of proton are shown for velocity of (a) and (d) 0.5 at.\ u., (b) and (e) 1.5 at.\ u., (c) and (f) 4.0 at.\ u.
}
\end{figure}

In the following, we provide deeper insight into the intricate dynamics of a proton on a [100] channel in In$_{0.5}$Ga$_{0.5}$P.
In particular, we disentangle the influence of electronic excitations on the dynamics by comparing Ehrenfest to BOMD.
To this end, Fig.\ \ref{fig:Dyns_BOMD_vs_ED} shows both the forces and the resulting displacement of the proton as it travels through the material for three different velocities.
The force acting on the projectile in different locations in the material in BOMD simulations does not depend on the projectile velocity, as confirmed by the solid lines in Fig.\ \ref{fig:Dyns_BOMD_vs_ED}.
This changes in Ehrenfest dynamics, and in the following we discuss the three different components of that force (see  Fig.\ \ref{fig:superlattice} for definition of $a$, $b$, and $c$).

Most notably, the force component parallel to the $a$ direction differs strongly between Ehrenfest and BOMD simulations.
This difference is completely expected and corresponds to electronic stopping, as discussed above.
As such, it is entirely attributed to non-adiabatic excitations that are captured by Ehrenfest dynamics, but not by BOMD simulations, for which the oscillations around zero force integrate to zero.

As shown in Fig.\ \ref{fig:Dyns_BOMD_vs_ED}, we also find non-zero forces for the $b$ and $c$ direction, but only non-zero displacement for $b$ direction within BOMD simulations.
The initial position of the proton at the center of the channel is not the equilibrium position in $b$ direction since In$_{0.5}$Ga$_{0.5}$P breaks the symmetry along the $b$ axis of InP and GaP.
As the proton moves through the material, it interacts with first-nearest-lattice atoms that repeat in the order In, P, Ga, and P in the directions $b$, $c$, $-b$, and $-c$, respectively.
It experiences repulsion from all of these atoms, but only the repulsion from P is oscillatory around zero.
In $b$ direction, the repulsion from In is larger than that from Ga, resulting in the displacement shown in Fig.\ \ref{fig:Dyns_BOMD_vs_ED}.

Furthermore, Fig.\ \ref{fig:Dyns_BOMD_vs_ED} illustrates that these force components acting on the proton parallel to the $b$ and $c$ directions become significantly larger and depend on the proton velocity within Ehrenfest dynamics.
While the overall shape of the force parallel to $b$ still strongly resembles the BOMD force, it becomes slightly more asymmetric, leading to larger displacements of the proton along this direction (see Fig.\ \ref{fig:Dyns_BOMD_vs_ED}).
More importantly, the force along $c$ significantly deviates from the BOMD force, both qualitatively and quantitatively, and even shows a different frequency of the oscillatory behavior.
Since these oscillations are not entirely symmetric around zero force, this leads to velocity-dependent displacements of the proton along $c$ that are absent in BOMD simulations (see Fig.\ \ref{fig:Dyns_BOMD_vs_ED}, in particular for $v$=1.5 at.\ u.).

Limited by the computational cost of Ehrenfest dynamics, we only report a trajectory of about 30 $a_\mathrm{B}$.
However, even for this short trajectory we clearly identify an effect of electronic excitations on the trajectory of the proton projectile.
While BOMD predicts deviations from an ideal trajectory along the center of the [100] lattice channel in In$_{0.5}$Ga$_{0.5}$P, this is amplified and becomes velocity-dependent in Ehrenfest dynamics, due to the excitation of electrons.
Velocity-dependent non-adiabatic forces caused by electronic excitations were identified before using RT-TDDFT \cite{Correa:2012}. 

Oscillations of projectiles have been reported before for channeling, however, the effect of electronic excitations is generally neglected \cite{Gemmell:1974}.
In this work, we accurately quantify this effect and while we find that the magnitude is small, our first-principles results provide the first direct quantitative evidence of an electronic contribution to such oscillations.
In particular, we show that these excited-electron contributions cause non-zero forces even for cases where BOMD finds zero force and, thus, significantly affect the dynamics of fast protons as it moves through the material.
More computational work and, ideally, longer Ehrenfest trajectories are necessary to further study this behavior.

\section{\label{sec:conclusion}Conclusions}

We reported on RT-TDDFT first-principles simulations to investigate  electronic stopping of protons in InP, GaP, and the CuAu-I ordered phase of In$_{0.5}$Ga$_{0.5}$P.
We compare our results from this parameter-free approach to data that we obtained using SRIM and find very good agreement for proton kinetic energies below about 25 keV.
The agreement is worse for higher kinetic energies, potentially due to core electronic states that were not treated as valence electrons in our pseudopotential implementation.
Nevertheless, we find a pronounced direction-dependence of electronic stopping along different channels and explain this using the magnitude of the electron density the proton projectile interacts with.
We also find a clear indication of local enhancement and reduction of stopping for the [100] channel, and explain this by local enhancement and reduction of the ground-state electron density.
The dependence of this effect on the proton velocity underscores its non-adiabatic character.

While these effects will be difficult to observe \emph{directly} in experiment, we conjecture that they significantly contribute to the dynamics of charged ions in semiconductor materials.
To investigate this further, we directly study the dynamics of a proton moving through In$_{0.5}$Ga$_{0.5}$P, using Ehrenfest and BOMD.
This comparison reveals an influence of electronic excitations both on force and displacement of the proton.
Even though the trajectories reported here are very short, they nevertheless illustrate that excited electronic states can trigger dynamics that is absent in a solid in its ground state.
We believe that these effects contribute to oscillations of charged projectiles as they move through a material.
Excited electronic states need to be taken into account in order to understand radiation damage on an atomistic level, and the use of TDDFT in an Ehrenfest MD scheme is a particularly appealing approach to do so, striking a desirable balance between accuracy and computational cost.

\begin{acknowledgments}
C.-W.\ L.\ acknowledges support from the Government Scholarship to Study Abroad from the Taiwan Ministry of Education.
Financial support from Sandia National Laboratories through the Sandia-UIUC collaboration is gratefully acknowledged (SNL grant no.\ 1736375).
An award of computer time was provided by the Innovative and Novel Computational Impact on Theory and Experiment (INCITE) program.
This research used resources of the Argonne Leadership Computing Facility, which is a DOE Office of Science User Facility supported under Contract DE-AC02-06CH11357. Data used in this work are available at the Materials Data Facility at http://dx.doi.org/10.18126/M2R337.
\end{acknowledgments}

\section*{Author contribution statement}

A.\ S.\ conceived the original idea and supervised the project. A.\ S.\ and C.-W.\ L.\ planned the simulations and C.-W.\ L.\ carried out all simulations. A.\ S.\ and C.-W.\ L.\ analyzed and discussed the results and both contributed to the manuscript.

\bibliography{InGaP.bib}

\begin{thebibliography}{64}%
\makeatletter
\providecommand \@ifxundefined [1]{%
 \@ifx{#1\undefined}
}%
\providecommand \@ifnum [1]{%
 \ifnum #1\expandafter \@firstoftwo
 \else \expandafter \@secondoftwo
 \fi
}%
\providecommand \@ifx [1]{%
 \ifx #1\expandafter \@firstoftwo
 \else \expandafter \@secondoftwo
 \fi
}%
\providecommand \natexlab [1]{#1}%
\providecommand \enquote  [1]{``#1''}%
\providecommand \bibnamefont  [1]{#1}%
\providecommand \bibfnamefont [1]{#1}%
\providecommand \citenamefont [1]{#1}%
\providecommand \href@noop [0]{\@secondoftwo}%
\providecommand \href [0]{\begingroup \@sanitize@url \@href}%
\providecommand \@href[1]{\@@startlink{#1}\@@href}%
\providecommand \@@href[1]{\endgroup#1\@@endlink}%
\providecommand \@sanitize@url [0]{\catcode `\\12\catcode `\$12\catcode
  `\&12\catcode `\#12\catcode `\^12\catcode `\_12\catcode `\%12\relax}%
\providecommand \@@startlink[1]{}%
\providecommand \@@endlink[0]{}%
\providecommand \url  [0]{\begingroup\@sanitize@url \@url }%
\providecommand \@url [1]{\endgroup\@href {#1}{\urlprefix }}%
\providecommand \urlprefix  [0]{URL }%
\providecommand \Eprint [0]{\href }%
\providecommand \doibase [0]{http://dx.doi.org/}%
\providecommand \selectlanguage [0]{\@gobble}%
\providecommand \bibinfo  [0]{\@secondoftwo}%
\providecommand \bibfield  [0]{\@secondoftwo}%
\providecommand \translation [1]{[#1]}%
\providecommand \BibitemOpen [0]{}%
\providecommand \bibitemStop [0]{}%
\providecommand \bibitemNoStop [0]{.\EOS\space}%
\providecommand \EOS [0]{\spacefactor3000\relax}%
\providecommand \BibitemShut  [1]{\csname bibitem#1\endcsname}%
\let\auto@bib@innerbib\@empty
%</preamble>
\bibitem [{\citenamefont {Pavesi}\ \emph {et~al.}(1991)\citenamefont {Pavesi},
  \citenamefont {Piazza}, \citenamefont {Rudra}, \citenamefont {Carlin},\ and\
  \citenamefont {Ilegems}}]{Pavesi:1991}%
  \BibitemOpen
  \bibfield  {author} {\bibinfo {author} {\bibfnamefont {L.}~\bibnamefont
  {Pavesi}}, \bibinfo {author} {\bibfnamefont {F.}~\bibnamefont {Piazza}},
  \bibinfo {author} {\bibfnamefont {A.}~\bibnamefont {Rudra}}, \bibinfo
  {author} {\bibfnamefont {J.~F.}\ \bibnamefont {Carlin}}, \ and\ \bibinfo
  {author} {\bibfnamefont {M.}~\bibnamefont {Ilegems}},\ }\href {\doibase
  10.1103/PhysRevB.44.9052} {\bibfield  {journal} {\bibinfo  {journal} {Phys.
  Rev. B}\ }\textbf {\bibinfo {volume} {44}},\ \bibinfo {pages} {9052}
  (\bibinfo {year} {1991})}\BibitemShut {NoStop}%
\bibitem [{\citenamefont {Bugajski}\ \emph {et~al.}(1983)\citenamefont
  {Bugajski}, \citenamefont {Kontkiewicz},\ and\ \citenamefont
  {Mariette}}]{Bugajski1983}%
  \BibitemOpen
  \bibfield  {author} {\bibinfo {author} {\bibfnamefont {M.}~\bibnamefont
  {Bugajski}}, \bibinfo {author} {\bibfnamefont {A.~M.}\ \bibnamefont
  {Kontkiewicz}}, \ and\ \bibinfo {author} {\bibfnamefont {H.}~\bibnamefont
  {Mariette}},\ }\href {\doibase 10.1103/PhysRevB.28.7105} {\bibfield
  {journal} {\bibinfo  {journal} {Phys. Rev. B}\ }\textbf {\bibinfo {volume}
  {28}},\ \bibinfo {pages} {7105} (\bibinfo {year} {1983})}\BibitemShut
  {NoStop}%
\bibitem [{\citenamefont {Panish}\ and\ \citenamefont
  {Jr.}(1969)}]{Panish1969}%
  \BibitemOpen
  \bibfield  {author} {\bibinfo {author} {\bibfnamefont {M.~B.}\ \bibnamefont
  {Panish}}\ and\ \bibinfo {author} {\bibfnamefont {H.~C.~C.}\ \bibnamefont
  {Jr.}},\ }\href {\doibase 10.1063/1.1657024} {\bibfield  {journal} {\bibinfo
  {journal} {J. Appl. Phys.}\ }\textbf {\bibinfo {volume} {40}},\ \bibinfo
  {pages} {163} (\bibinfo {year} {1969})}\BibitemShut {NoStop}%
\bibitem [{\citenamefont {Takamoto}\ \emph {et~al.}(1997)\citenamefont
  {Takamoto}, \citenamefont {Ikeda}, \citenamefont {Kurita},\ and\
  \citenamefont {Ohmori}}]{Takamoto1997}%
  \BibitemOpen
  \bibfield  {author} {\bibinfo {author} {\bibfnamefont {T.}~\bibnamefont
  {Takamoto}}, \bibinfo {author} {\bibfnamefont {E.}~\bibnamefont {Ikeda}},
  \bibinfo {author} {\bibfnamefont {H.}~\bibnamefont {Kurita}}, \ and\ \bibinfo
  {author} {\bibfnamefont {M.}~\bibnamefont {Ohmori}},\ }\href {\doibase
  10.1063/1.118419} {\bibfield  {journal} {\bibinfo  {journal} {Appl. Phys.
  Lett.}\ }\textbf {\bibinfo {volume} {70}},\ \bibinfo {pages} {381} (\bibinfo
  {year} {1997})}\BibitemShut {NoStop}%
\bibitem [{\citenamefont {King}\ \emph {et~al.}(2007)\citenamefont {King},
  \citenamefont {Law}, \citenamefont {Edmondson}, \citenamefont {Fetzer},
  \citenamefont {Kinsey}, \citenamefont {Yoon}, \citenamefont {Sherif},\ and\
  \citenamefont {Karam}}]{King2007}%
  \BibitemOpen
  \bibfield  {author} {\bibinfo {author} {\bibfnamefont {R.~R.}\ \bibnamefont
  {King}}, \bibinfo {author} {\bibfnamefont {D.~C.}\ \bibnamefont {Law}},
  \bibinfo {author} {\bibfnamefont {K.~M.}\ \bibnamefont {Edmondson}}, \bibinfo
  {author} {\bibfnamefont {C.~M.}\ \bibnamefont {Fetzer}}, \bibinfo {author}
  {\bibfnamefont {G.~S.}\ \bibnamefont {Kinsey}}, \bibinfo {author}
  {\bibfnamefont {H.}~\bibnamefont {Yoon}}, \bibinfo {author} {\bibfnamefont
  {R.~A.}\ \bibnamefont {Sherif}}, \ and\ \bibinfo {author} {\bibfnamefont
  {N.~H.}\ \bibnamefont {Karam}},\ }\href {\doibase 10.1063/1.2734507}
  {\bibfield  {journal} {\bibinfo  {journal} {Appl. Phys. Lett.}\ }\textbf
  {\bibinfo {volume} {90}},\ \bibinfo {pages} {183516} (\bibinfo {year}
  {2007})}\BibitemShut {NoStop}%
\bibitem [{\citenamefont {Yamaguchi}(2001)}]{YAMAGUCHI2001}%
  \BibitemOpen
  \bibfield  {author} {\bibinfo {author} {\bibfnamefont {M.}~\bibnamefont
  {Yamaguchi}},\ }\href {\doibase
  https://doi.org/10.1016/S0927-0248(00)00344-5} {\bibfield  {journal}
  {\bibinfo  {journal} {Sol. Energ. Mat. Sol. C.}\ }\textbf {\bibinfo {volume}
  {68}},\ \bibinfo {pages} {31 } (\bibinfo {year} {2001})},\ \bibinfo {note}
  {solar cells in space}\BibitemShut {NoStop}%
\bibitem [{\citenamefont {Dharmarasu}\ \emph {et~al.}(2001)\citenamefont
  {Dharmarasu}, \citenamefont {Yamaguchi}, \citenamefont {Khan}, \citenamefont
  {Yamada}, \citenamefont {Tanabe}, \citenamefont {Takagishi}, \citenamefont
  {Takamoto}, \citenamefont {Ohshima}, \citenamefont {Itoh}, \citenamefont
  {Imaizumi},\ and\ \citenamefont {Matsuda}}]{Dharmarasu2001}%
  \BibitemOpen
  \bibfield  {author} {\bibinfo {author} {\bibfnamefont {N.}~\bibnamefont
  {Dharmarasu}}, \bibinfo {author} {\bibfnamefont {M.}~\bibnamefont
  {Yamaguchi}}, \bibinfo {author} {\bibfnamefont {A.}~\bibnamefont {Khan}},
  \bibinfo {author} {\bibfnamefont {T.}~\bibnamefont {Yamada}}, \bibinfo
  {author} {\bibfnamefont {T.}~\bibnamefont {Tanabe}}, \bibinfo {author}
  {\bibfnamefont {S.}~\bibnamefont {Takagishi}}, \bibinfo {author}
  {\bibfnamefont {T.}~\bibnamefont {Takamoto}}, \bibinfo {author}
  {\bibfnamefont {T.}~\bibnamefont {Ohshima}}, \bibinfo {author} {\bibfnamefont
  {H.}~\bibnamefont {Itoh}}, \bibinfo {author} {\bibfnamefont {M.}~\bibnamefont
  {Imaizumi}}, \ and\ \bibinfo {author} {\bibfnamefont {S.}~\bibnamefont
  {Matsuda}},\ }\href {\doibase 10.1063/1.1409270} {\bibfield  {journal}
  {\bibinfo  {journal} {Appl. Phys. Lett.}\ }\textbf {\bibinfo {volume} {79}},\
  \bibinfo {pages} {2399} (\bibinfo {year} {2001})}\BibitemShut {NoStop}%
\bibitem [{\citenamefont {Cress}\ \emph {et~al.}(2006)\citenamefont {Cress},
  \citenamefont {Landi}, \citenamefont {Raffaelle},\ and\ \citenamefont
  {Wilt}}]{Cress2006}%
  \BibitemOpen
  \bibfield  {author} {\bibinfo {author} {\bibfnamefont {C.~D.}\ \bibnamefont
  {Cress}}, \bibinfo {author} {\bibfnamefont {B.~J.}\ \bibnamefont {Landi}},
  \bibinfo {author} {\bibfnamefont {R.~P.}\ \bibnamefont {Raffaelle}}, \ and\
  \bibinfo {author} {\bibfnamefont {D.~M.}\ \bibnamefont {Wilt}},\ }\href
  {\doibase 10.1063/1.2390623} {\bibfield  {journal} {\bibinfo  {journal} {J.
  Appl. Phys.}\ }\textbf {\bibinfo {volume} {100}},\ \bibinfo {pages} {114519}
  (\bibinfo {year} {2006})}\BibitemShut {NoStop}%
\bibitem [{\citenamefont {Yamaguchi}\ \emph {et~al.}(1984)\citenamefont
  {Yamaguchi}, \citenamefont {Uemura},\ and\ \citenamefont
  {Yamamoto}}]{Yamaguchi1984}%
  \BibitemOpen
  \bibfield  {author} {\bibinfo {author} {\bibfnamefont {M.}~\bibnamefont
  {Yamaguchi}}, \bibinfo {author} {\bibfnamefont {C.}~\bibnamefont {Uemura}}, \
  and\ \bibinfo {author} {\bibfnamefont {A.}~\bibnamefont {Yamamoto}},\ }\href
  {\doibase 10.1063/1.333396} {\bibfield  {journal} {\bibinfo  {journal} {J.
  Appl. Phys.}\ }\textbf {\bibinfo {volume} {55}},\ \bibinfo {pages} {1429}
  (\bibinfo {year} {1984})}\BibitemShut {NoStop}%
\bibitem [{\citenamefont {Yamaguchi}\ \emph {et~al.}(1997)\citenamefont
  {Yamaguchi}, \citenamefont {Okuda}, \citenamefont {Taylor}, \citenamefont
  {Takamoto}, \citenamefont {Ikeda},\ and\ \citenamefont
  {Kurita}}]{Yamaguchi1997}%
  \BibitemOpen
  \bibfield  {author} {\bibinfo {author} {\bibfnamefont {M.}~\bibnamefont
  {Yamaguchi}}, \bibinfo {author} {\bibfnamefont {T.}~\bibnamefont {Okuda}},
  \bibinfo {author} {\bibfnamefont {S.~J.}\ \bibnamefont {Taylor}}, \bibinfo
  {author} {\bibfnamefont {T.}~\bibnamefont {Takamoto}}, \bibinfo {author}
  {\bibfnamefont {E.}~\bibnamefont {Ikeda}}, \ and\ \bibinfo {author}
  {\bibfnamefont {H.}~\bibnamefont {Kurita}},\ }\href {\doibase
  10.1063/1.118618} {\bibfield  {journal} {\bibinfo  {journal} {Appl. Phys.
  Lett.}\ }\textbf {\bibinfo {volume} {70}},\ \bibinfo {pages} {1566} (\bibinfo
  {year} {1997})}\BibitemShut {NoStop}%
\bibitem [{\citenamefont {Shockley}\ and\ \citenamefont
  {Read}(1952)}]{Shockley1952}%
  \BibitemOpen
  \bibfield  {author} {\bibinfo {author} {\bibfnamefont {W.}~\bibnamefont
  {Shockley}}\ and\ \bibinfo {author} {\bibfnamefont {W.~T.}\ \bibnamefont
  {Read}},\ }\href {\doibase 10.1103/PhysRev.87.835} {\bibfield  {journal}
  {\bibinfo  {journal} {Phys. Rev.}\ }\textbf {\bibinfo {volume} {87}},\
  \bibinfo {pages} {835} (\bibinfo {year} {1952})}\BibitemShut {NoStop}%
\bibitem [{\citenamefont {Hall}(1952)}]{Hall1952}%
  \BibitemOpen
  \bibfield  {author} {\bibinfo {author} {\bibfnamefont {R.~N.}\ \bibnamefont
  {Hall}},\ }\href {\doibase 10.1103/PhysRev.87.387} {\bibfield  {journal}
  {\bibinfo  {journal} {Phys. Rev.}\ }\textbf {\bibinfo {volume} {87}},\
  \bibinfo {pages} {387} (\bibinfo {year} {1952})}\BibitemShut {NoStop}%
\bibitem [{\citenamefont {Yamaguchi}(1995)}]{Yamaguchi1995}%
  \BibitemOpen
  \bibfield  {author} {\bibinfo {author} {\bibfnamefont {M.}~\bibnamefont
  {Yamaguchi}},\ }\href {\doibase 10.1063/1.360236} {\bibfield  {journal}
  {\bibinfo  {journal} {J. Appl. Phys.}\ }\textbf {\bibinfo {volume} {78}},\
  \bibinfo {pages} {1476} (\bibinfo {year} {1995})}\BibitemShut {NoStop}%
\bibitem [{\citenamefont {Bourgoin}\ and\ \citenamefont
  {Corbett}(1972)}]{BOURGOIN1972}%
  \BibitemOpen
  \bibfield  {author} {\bibinfo {author} {\bibfnamefont {J.}~\bibnamefont
  {Bourgoin}}\ and\ \bibinfo {author} {\bibfnamefont {J.}~\bibnamefont
  {Corbett}},\ }\href {\doibase https://doi.org/10.1016/0375-9601(72)90523-3}
  {\bibfield  {journal} {\bibinfo  {journal} {Phys. Lett. A}\ }\textbf
  {\bibinfo {volume} {38}},\ \bibinfo {pages} {135 } (\bibinfo {year}
  {1972})}\BibitemShut {NoStop}%
\bibitem [{\citenamefont {Itoh}(1998)}]{ITOH1998}%
  \BibitemOpen
  \bibfield  {author} {\bibinfo {author} {\bibfnamefont {N.}~\bibnamefont
  {Itoh}},\ }\href {\doibase https://doi.org/10.1016/S0168-583X(97)00523-5}
  {\bibfield  {journal} {\bibinfo  {journal} {Nucl. Instrum. Meth. B}\ }\textbf
  {\bibinfo {volume} {135}},\ \bibinfo {pages} {175} (\bibinfo {year}
  {1998})}\BibitemShut {NoStop}%
\bibitem [{\citenamefont {Bai}\ \emph {et~al.}(2010)\citenamefont {Bai},
  \citenamefont {Voter}, \citenamefont {Hoagland}, \citenamefont {Nastasi},\
  and\ \citenamefont {Uberuaga}}]{Bai2010}%
  \BibitemOpen
  \bibfield  {author} {\bibinfo {author} {\bibfnamefont {X.-M.}\ \bibnamefont
  {Bai}}, \bibinfo {author} {\bibfnamefont {A.~F.}\ \bibnamefont {Voter}},
  \bibinfo {author} {\bibfnamefont {R.~G.}\ \bibnamefont {Hoagland}}, \bibinfo
  {author} {\bibfnamefont {M.}~\bibnamefont {Nastasi}}, \ and\ \bibinfo
  {author} {\bibfnamefont {B.~P.}\ \bibnamefont {Uberuaga}},\ }\href {\doibase
  10.1126/science.1183723} {\bibfield  {journal} {\bibinfo  {journal}
  {Science}\ }\textbf {\bibinfo {volume} {327}},\ \bibinfo {pages} {1631}
  (\bibinfo {year} {2010})}\BibitemShut {NoStop}%
\bibitem [{\citenamefont {Klatt}\ \emph {et~al.}(1993)\citenamefont {Klatt},
  \citenamefont {Averback}, \citenamefont {Forbes},\ and\ \citenamefont
  {Coleman}}]{Klatt1993}%
  \BibitemOpen
  \bibfield  {author} {\bibinfo {author} {\bibfnamefont {J.~L.}\ \bibnamefont
  {Klatt}}, \bibinfo {author} {\bibfnamefont {R.~S.}\ \bibnamefont {Averback}},
  \bibinfo {author} {\bibfnamefont {D.~V.}\ \bibnamefont {Forbes}}, \ and\
  \bibinfo {author} {\bibfnamefont {J.~J.}\ \bibnamefont {Coleman}},\ }\href
  {\doibase 10.1103/PhysRevB.48.17629} {\bibfield  {journal} {\bibinfo
  {journal} {Phys. Rev. B}\ }\textbf {\bibinfo {volume} {48}},\ \bibinfo
  {pages} {17629} (\bibinfo {year} {1993})}\BibitemShut {NoStop}%
\bibitem [{\citenamefont {Jiang}\ \emph {et~al.}(2018)\citenamefont {Jiang},
  \citenamefont {Xiao}, \citenamefont {Peng}, \citenamefont {Yang},
  \citenamefont {Liu},\ and\ \citenamefont {Zu}}]{Jiang2018}%
  \BibitemOpen
  \bibfield  {author} {\bibinfo {author} {\bibfnamefont {M.}~\bibnamefont
  {Jiang}}, \bibinfo {author} {\bibfnamefont {H.~Y.}\ \bibnamefont {Xiao}},
  \bibinfo {author} {\bibfnamefont {S.~M.}\ \bibnamefont {Peng}}, \bibinfo
  {author} {\bibfnamefont {G.~X.}\ \bibnamefont {Yang}}, \bibinfo {author}
  {\bibfnamefont {Z.~J.}\ \bibnamefont {Liu}}, \ and\ \bibinfo {author}
  {\bibfnamefont {X.~T.}\ \bibnamefont {Zu}},\ }\href {\doibase
  10.1038/s41598-018-20155-0} {\bibfield  {journal} {\bibinfo  {journal} {Sci.
  Rep.-UK}\ }\textbf {\bibinfo {volume} {8}},\ \bibinfo {pages} {2012}
  (\bibinfo {year} {2018})}\BibitemShut {NoStop}%
\bibitem [{\citenamefont {Botti}\ \emph {et~al.}(2002)\citenamefont {Botti},
  \citenamefont {Vast}, \citenamefont {Reining}, \citenamefont {Olevano},\ and\
  \citenamefont {Andreani}}]{Botti2002}%
  \BibitemOpen
  \bibfield  {author} {\bibinfo {author} {\bibfnamefont {S.}~\bibnamefont
  {Botti}}, \bibinfo {author} {\bibfnamefont {N.}~\bibnamefont {Vast}},
  \bibinfo {author} {\bibfnamefont {L.}~\bibnamefont {Reining}}, \bibinfo
  {author} {\bibfnamefont {V.}~\bibnamefont {Olevano}}, \ and\ \bibinfo
  {author} {\bibfnamefont {L.~C.}\ \bibnamefont {Andreani}},\ }\href {\doibase
  10.1103/PhysRevLett.89.216803} {\bibfield  {journal} {\bibinfo  {journal}
  {Phys. Rev. Lett.}\ }\textbf {\bibinfo {volume} {89}},\ \bibinfo {pages}
  {216803} (\bibinfo {year} {2002})}\BibitemShut {NoStop}%
\bibitem [{\citenamefont {Botti}\ \emph {et~al.}(2004)\citenamefont {Botti},
  \citenamefont {Vast}, \citenamefont {Reining}, \citenamefont {Olevano},\ and\
  \citenamefont {Andreani}}]{Botti2004}%
  \BibitemOpen
  \bibfield  {author} {\bibinfo {author} {\bibfnamefont {S.}~\bibnamefont
  {Botti}}, \bibinfo {author} {\bibfnamefont {N.}~\bibnamefont {Vast}},
  \bibinfo {author} {\bibfnamefont {L.}~\bibnamefont {Reining}}, \bibinfo
  {author} {\bibfnamefont {V.}~\bibnamefont {Olevano}}, \ and\ \bibinfo
  {author} {\bibfnamefont {L.~C.}\ \bibnamefont {Andreani}},\ }\href {\doibase
  10.1103/PhysRevB.70.045301} {\bibfield  {journal} {\bibinfo  {journal} {Phys.
  Rev. B}\ }\textbf {\bibinfo {volume} {70}},\ \bibinfo {pages} {045301}
  (\bibinfo {year} {2004})}\BibitemShut {NoStop}%
\bibitem [{\citenamefont {Gumbs}(1988)}]{Godfrey1988}%
  \BibitemOpen
  \bibfield  {author} {\bibinfo {author} {\bibfnamefont {G.}~\bibnamefont
  {Gumbs}},\ }\href {\doibase 10.1103/PhysRevB.37.10184} {\bibfield  {journal}
  {\bibinfo  {journal} {Phys. Rev. B}\ }\textbf {\bibinfo {volume} {37}},\
  \bibinfo {pages} {10184} (\bibinfo {year} {1988})}\BibitemShut {NoStop}%
\bibitem [{\citenamefont {Bethe}(1930)}]{Bethe1930}%
  \BibitemOpen
  \bibfield  {author} {\bibinfo {author} {\bibfnamefont {H.}~\bibnamefont
  {Bethe}},\ }\href {\doibase 10.1002/andp.19303970303} {\bibfield  {journal}
  {\bibinfo  {journal} {Ann. Phys.}\ }\textbf {\bibinfo {volume} {397}},\
  \bibinfo {pages} {325} (\bibinfo {year} {1930})}\BibitemShut {NoStop}%
\bibitem [{\citenamefont {Cruz}(2012)}]{Salvador2012}%
  \BibitemOpen
  \bibfield  {author} {\bibinfo {author} {\bibfnamefont {S.~A.}\ \bibnamefont
  {Cruz}},\ }\href {\doibase 10.1080/10420150.2011.642873} {\bibfield
  {journal} {\bibinfo  {journal} {Radia. Eff. Defect. S.}\ }\textbf {\bibinfo
  {volume} {167}},\ \bibinfo {pages} {621} (\bibinfo {year}
  {2012})}\BibitemShut {NoStop}%
\bibitem [{\citenamefont {Stringfellow}\ and\ \citenamefont
  {Chen}(1991)}]{Stringfellow1991}%
  \BibitemOpen
  \bibfield  {author} {\bibinfo {author} {\bibfnamefont {G.~B.}\ \bibnamefont
  {Stringfellow}}\ and\ \bibinfo {author} {\bibfnamefont {G.~S.}\ \bibnamefont
  {Chen}},\ }\href {\doibase 10.1116/1.585761} {\bibfield  {journal} {\bibinfo
  {journal} {J. Vac. Sci. Technol. B}\ }\textbf {\bibinfo {volume} {9}},\
  \bibinfo {pages} {2182} (\bibinfo {year} {1991})}\BibitemShut {NoStop}%
\bibitem [{\citenamefont {Kuan}\ \emph {et~al.}(1985)\citenamefont {Kuan},
  \citenamefont {Kuech}, \citenamefont {Wang},\ and\ \citenamefont
  {Wilkie}}]{Kuan1985}%
  \BibitemOpen
  \bibfield  {author} {\bibinfo {author} {\bibfnamefont {T.~S.}\ \bibnamefont
  {Kuan}}, \bibinfo {author} {\bibfnamefont {T.~F.}\ \bibnamefont {Kuech}},
  \bibinfo {author} {\bibfnamefont {W.~I.}\ \bibnamefont {Wang}}, \ and\
  \bibinfo {author} {\bibfnamefont {E.~L.}\ \bibnamefont {Wilkie}},\ }\href
  {\doibase 10.1103/PhysRevLett.54.201} {\bibfield  {journal} {\bibinfo
  {journal} {Phys. Rev. Lett.}\ }\textbf {\bibinfo {volume} {54}},\ \bibinfo
  {pages} {201} (\bibinfo {year} {1985})}\BibitemShut {NoStop}%
\bibitem [{\citenamefont {Suzuki}\ \emph {et~al.}(1988)\citenamefont {Suzuki},
  \citenamefont {Gomyo}, \citenamefont {Iijima}, \citenamefont {Kobayashi},
  \citenamefont {Kawata}, \citenamefont {Hino},\ and\ \citenamefont
  {Yuasa}}]{Suzuki1988}%
  \BibitemOpen
  \bibfield  {author} {\bibinfo {author} {\bibfnamefont {T.}~\bibnamefont
  {Suzuki}}, \bibinfo {author} {\bibfnamefont {A.}~\bibnamefont {Gomyo}},
  \bibinfo {author} {\bibfnamefont {S.}~\bibnamefont {Iijima}}, \bibinfo
  {author} {\bibfnamefont {K.}~\bibnamefont {Kobayashi}}, \bibinfo {author}
  {\bibfnamefont {S.}~\bibnamefont {Kawata}}, \bibinfo {author} {\bibfnamefont
  {I.}~\bibnamefont {Hino}}, \ and\ \bibinfo {author} {\bibfnamefont
  {T.}~\bibnamefont {Yuasa}},\ }\href {\doibase 10.1143/JJAP.27.2098}
  {\bibfield  {journal} {\bibinfo  {journal} {Jpn. J. Appl. Phys.}\ }\textbf
  {\bibinfo {volume} {27}},\ \bibinfo {pages} {2098} (\bibinfo {year}
  {1988})}\BibitemShut {NoStop}%
\bibitem [{\citenamefont {Wei}\ and\ \citenamefont {Zunger}(1994)}]{Wei1994}%
  \BibitemOpen
  \bibfield  {author} {\bibinfo {author} {\bibfnamefont {S.-H.}\ \bibnamefont
  {Wei}}\ and\ \bibinfo {author} {\bibfnamefont {A.}~\bibnamefont {Zunger}},\
  }\href {\doibase 10.1103/PhysRevB.49.14337} {\bibfield  {journal} {\bibinfo
  {journal} {Phys. Rev. B}\ }\textbf {\bibinfo {volume} {49}},\ \bibinfo
  {pages} {14337} (\bibinfo {year} {1994})}\BibitemShut {NoStop}%
\bibitem [{\citenamefont {Hassine}\ \emph {et~al.}(1996)\citenamefont
  {Hassine}, \citenamefont {Sapriel}, \citenamefont {Le~Berre}, \citenamefont
  {Di~Forte-Poisson}, \citenamefont {Alexandre},\ and\ \citenamefont
  {Quillec}}]{Hassine1996}%
  \BibitemOpen
  \bibfield  {author} {\bibinfo {author} {\bibfnamefont {A.}~\bibnamefont
  {Hassine}}, \bibinfo {author} {\bibfnamefont {J.}~\bibnamefont {Sapriel}},
  \bibinfo {author} {\bibfnamefont {P.}~\bibnamefont {Le~Berre}}, \bibinfo
  {author} {\bibfnamefont {M.~A.}\ \bibnamefont {Di~Forte-Poisson}}, \bibinfo
  {author} {\bibfnamefont {F.}~\bibnamefont {Alexandre}}, \ and\ \bibinfo
  {author} {\bibfnamefont {M.}~\bibnamefont {Quillec}},\ }\href {\doibase
  10.1103/PhysRevB.54.2728} {\bibfield  {journal} {\bibinfo  {journal} {Phys.
  Rev. B}\ }\textbf {\bibinfo {volume} {54}},\ \bibinfo {pages} {2728}
  (\bibinfo {year} {1996})}\BibitemShut {NoStop}%
\bibitem [{\citenamefont {Ozoli\ifmmode \mbox{\c{n}}\else
  \c{n}\fi{}\ifmmode~\check{s}\else \v{s}\fi{}}\ and\ \citenamefont
  {Zunger}(1998)}]{Ozolins1998}%
  \BibitemOpen
  \bibfield  {author} {\bibinfo {author} {\bibfnamefont {V.}~\bibnamefont
  {Ozoli\ifmmode \mbox{\c{n}}\else \c{n}\fi{}\ifmmode~\check{s}\else
  \v{s}\fi{}}}\ and\ \bibinfo {author} {\bibfnamefont {A.}~\bibnamefont
  {Zunger}},\ }\href {\doibase 10.1103/PhysRevB.57.R9404} {\bibfield  {journal}
  {\bibinfo  {journal} {Phys. Rev. B}\ }\textbf {\bibinfo {volume} {57}},\
  \bibinfo {pages} {R9404} (\bibinfo {year} {1998})}\BibitemShut {NoStop}%
\bibitem [{\citenamefont {Duda}\ \emph {et~al.}(2011)\citenamefont {Duda},
  \citenamefont {English}, \citenamefont {Jordan}, \citenamefont {Norris},\
  and\ \citenamefont {Soffa}}]{Duda2011}%
  \BibitemOpen
  \bibfield  {author} {\bibinfo {author} {\bibfnamefont {J.~C.}\ \bibnamefont
  {Duda}}, \bibinfo {author} {\bibfnamefont {T.~S.}\ \bibnamefont {English}},
  \bibinfo {author} {\bibfnamefont {D.~A.}\ \bibnamefont {Jordan}}, \bibinfo
  {author} {\bibfnamefont {P.~M.}\ \bibnamefont {Norris}}, \ and\ \bibinfo
  {author} {\bibfnamefont {W.~A.}\ \bibnamefont {Soffa}},\ }\href {\doibase
  10.1088/0953-8984/23/20/205401} {\bibfield  {journal} {\bibinfo  {journal}
  {J. Phys.-Condens. Mat.}\ }\textbf {\bibinfo {volume} {23}},\ \bibinfo
  {pages} {205401} (\bibinfo {year} {2011})}\BibitemShut {NoStop}%
\bibitem [{\citenamefont {Chernyak}\ \emph {et~al.}(1997)\citenamefont
  {Chernyak}, \citenamefont {Osinsky}, \citenamefont {Temkin}, \citenamefont
  {Mintairov}, \citenamefont {Malkina}, \citenamefont {Zvonkov},\ and\
  \citenamefont {Saf'anov}}]{Chernyak1997}%
  \BibitemOpen
  \bibfield  {author} {\bibinfo {author} {\bibfnamefont {L.}~\bibnamefont
  {Chernyak}}, \bibinfo {author} {\bibfnamefont {A.}~\bibnamefont {Osinsky}},
  \bibinfo {author} {\bibfnamefont {H.}~\bibnamefont {Temkin}}, \bibinfo
  {author} {\bibfnamefont {A.}~\bibnamefont {Mintairov}}, \bibinfo {author}
  {\bibfnamefont {I.~G.}\ \bibnamefont {Malkina}}, \bibinfo {author}
  {\bibfnamefont {B.~N.}\ \bibnamefont {Zvonkov}}, \ and\ \bibinfo {author}
  {\bibfnamefont {Y.~N.}\ \bibnamefont {Saf'anov}},\ }\href {\doibase
  10.1063/1.118864} {\bibfield  {journal} {\bibinfo  {journal} {Appl. Phys.
  Lett.}\ }\textbf {\bibinfo {volume} {70}},\ \bibinfo {pages} {2425} (\bibinfo
  {year} {1997})}\BibitemShut {NoStop}%
\bibitem [{\citenamefont {Gygi}()}]{qbox_davis}%
  \BibitemOpen
  \bibfield  {author} {\bibinfo {author} {\bibfnamefont {F.}~\bibnamefont
  {Gygi}},\ }\href@noop {} {\emph {\bibinfo {title} {Qbox open source code
  project}}},\ \bibinfo {type} {Tech. Rep.}\ (\bibinfo  {institution}
  {University of California, Davis})\ \bibinfo {note}
  {http://eslab.ucdavis.edu/}\BibitemShut {NoStop}%
\bibitem [{\citenamefont {Draeger}\ and\ \citenamefont
  {Gygi}(2017)}]{qball2017}%
  \BibitemOpen
  \bibfield  {author} {\bibinfo {author} {\bibfnamefont {E.~W.}\ \bibnamefont
  {Draeger}}\ and\ \bibinfo {author} {\bibfnamefont {F.}~\bibnamefont {Gygi}},\
  }\href {https://github.com/LLNL/qball} {\enquote {\bibinfo {title} {Qbox
  code, {Qb@ll} version},}\ } (\bibinfo {year} {2017}),\ \bibinfo {note}
  {{Lawrence Livermore National Laboratory}}\BibitemShut {NoStop}%
\bibitem [{\citenamefont {Hohenberg}\ and\ \citenamefont
  {Kohn}(1964)}]{Hohenberg1964}%
  \BibitemOpen
  \bibfield  {author} {\bibinfo {author} {\bibfnamefont {P.}~\bibnamefont
  {Hohenberg}}\ and\ \bibinfo {author} {\bibfnamefont {W.}~\bibnamefont
  {Kohn}},\ }\href {\doibase 10.1103/PhysRev.136.B864} {\bibfield  {journal}
  {\bibinfo  {journal} {Phys. Rev.}\ }\textbf {\bibinfo {volume} {136}},\
  \bibinfo {pages} {B864} (\bibinfo {year} {1964})}\BibitemShut {NoStop}%
\bibitem [{\citenamefont {Kohn}\ and\ \citenamefont {Sham}(1965)}]{Kohn1965}%
  \BibitemOpen
  \bibfield  {author} {\bibinfo {author} {\bibfnamefont {W.}~\bibnamefont
  {Kohn}}\ and\ \bibinfo {author} {\bibfnamefont {L.~J.}\ \bibnamefont
  {Sham}},\ }\href {\doibase 10.1103/PhysRev.140.A1133} {\bibfield  {journal}
  {\bibinfo  {journal} {Phys. Rev.}\ }\textbf {\bibinfo {volume} {140}},\
  \bibinfo {pages} {A1133} (\bibinfo {year} {1965})}\BibitemShut {NoStop}%
\bibitem [{\citenamefont {Kim}\ \emph {et~al.}(1999)\citenamefont {Kim},
  \citenamefont {Asahi}, \citenamefont {Asami},\ and\ \citenamefont {ichi
  Gonda}}]{Kim1999}%
  \BibitemOpen
  \bibfield  {author} {\bibinfo {author} {\bibfnamefont {S.-J.}\ \bibnamefont
  {Kim}}, \bibinfo {author} {\bibfnamefont {H.}~\bibnamefont {Asahi}}, \bibinfo
  {author} {\bibfnamefont {K.}~\bibnamefont {Asami}}, \ and\ \bibinfo {author}
  {\bibfnamefont {S.}~\bibnamefont {ichi Gonda}},\ }\href
  {http://stacks.iop.org/1347-4065/38/i=12A/a=L1372} {\bibfield  {journal}
  {\bibinfo  {journal} {Jpn. J. Appl. Phys.}\ }\textbf {\bibinfo {volume}
  {38}},\ \bibinfo {pages} {L1372} (\bibinfo {year} {1999})}\BibitemShut
  {NoStop}%
\bibitem [{\citenamefont {Ueda}\ \emph {et~al.}(1987)\citenamefont {Ueda},
  \citenamefont {Takikawa}, \citenamefont {Komeno},\ and\ \citenamefont
  {Umebu}}]{Ueda1987}%
  \BibitemOpen
  \bibfield  {author} {\bibinfo {author} {\bibfnamefont {O.}~\bibnamefont
  {Ueda}}, \bibinfo {author} {\bibfnamefont {M.}~\bibnamefont {Takikawa}},
  \bibinfo {author} {\bibfnamefont {J.}~\bibnamefont {Komeno}}, \ and\ \bibinfo
  {author} {\bibfnamefont {I.}~\bibnamefont {Umebu}},\ }\href
  {http://stacks.iop.org/1347-4065/26/i=11A/a=L1824} {\bibfield  {journal}
  {\bibinfo  {journal} {Jpn. J. Appl. Phys.}\ }\textbf {\bibinfo {volume}
  {26}},\ \bibinfo {pages} {L1824} (\bibinfo {year} {1987})}\BibitemShut
  {NoStop}%
\bibitem [{\citenamefont {Bellon}\ \emph {et~al.}(1988)\citenamefont {Bellon},
  \citenamefont {Chevalier}, \citenamefont {Martin}, \citenamefont
  {Dupont‐Nivet}, \citenamefont {Thiebaut},\ and\ \citenamefont
  {André}}]{Bellon1988}%
  \BibitemOpen
  \bibfield  {author} {\bibinfo {author} {\bibfnamefont {P.}~\bibnamefont
  {Bellon}}, \bibinfo {author} {\bibfnamefont {J.~P.}\ \bibnamefont
  {Chevalier}}, \bibinfo {author} {\bibfnamefont {G.~P.}\ \bibnamefont
  {Martin}}, \bibinfo {author} {\bibfnamefont {E.}~\bibnamefont
  {Dupont‐Nivet}}, \bibinfo {author} {\bibfnamefont {C.}~\bibnamefont
  {Thiebaut}}, \ and\ \bibinfo {author} {\bibfnamefont {J.~P.}\ \bibnamefont
  {André}},\ }\href {\doibase 10.1063/1.99419} {\bibfield  {journal} {\bibinfo
   {journal} {Appl. Phys. Lett.}\ }\textbf {\bibinfo {volume} {52}},\ \bibinfo
  {pages} {567} (\bibinfo {year} {1988})}\BibitemShut {NoStop}%
\bibitem [{\citenamefont {Ceperley}\ and\ \citenamefont
  {Alder}(1980)}]{Ceperley:1980}%
  \BibitemOpen
  \bibfield  {author} {\bibinfo {author} {\bibfnamefont {D.~M.}\ \bibnamefont
  {Ceperley}}\ and\ \bibinfo {author} {\bibfnamefont {B.~J.}\ \bibnamefont
  {Alder}},\ }\href {\doibase 10.1103/PhysRevLett.45.566} {\bibfield  {journal}
  {\bibinfo  {journal} {Phys. Rev. Lett.}\ }\textbf {\bibinfo {volume} {45}},\
  \bibinfo {pages} {566} (\bibinfo {year} {1980})}\BibitemShut {NoStop}%
\bibitem [{\citenamefont {Perdew}\ and\ \citenamefont
  {Zunger}(1981)}]{Perdew:1981}%
  \BibitemOpen
  \bibfield  {author} {\bibinfo {author} {\bibfnamefont {J.~P.}\ \bibnamefont
  {Perdew}}\ and\ \bibinfo {author} {\bibfnamefont {A.}~\bibnamefont
  {Zunger}},\ }\href {\doibase 10.1103/PhysRevB.23.5048} {\bibfield  {journal}
  {\bibinfo  {journal} {Phys. Rev. B}\ }\textbf {\bibinfo {volume} {23}},\
  \bibinfo {pages} {5048} (\bibinfo {year} {1981})}\BibitemShut {NoStop}%
\bibitem [{\citenamefont {Vanderbilt}(1985)}]{Vanderbilt:1985}%
  \BibitemOpen
  \bibfield  {author} {\bibinfo {author} {\bibfnamefont {D.}~\bibnamefont
  {Vanderbilt}},\ }\href {\doibase 10.1103/PhysRevB.32.8412} {\bibfield
  {journal} {\bibinfo  {journal} {Phys. Rev. B}\ }\textbf {\bibinfo {volume}
  {32}},\ \bibinfo {pages} {8412} (\bibinfo {year} {1985})}\BibitemShut
  {NoStop}%
\bibitem [{\citenamefont {Murnaghan}(1944)}]{Murnaghan1944}%
  \BibitemOpen
  \bibfield  {author} {\bibinfo {author} {\bibfnamefont {F.~D.}\ \bibnamefont
  {Murnaghan}},\ }\href@noop {} {\enquote {\bibinfo {title} {The
  compressibility of media under extreme pressures},}\ } (\bibinfo {year}
  {1944})\BibitemShut {NoStop}%
\bibitem [{\citenamefont {Marx}\ and\ \citenamefont {Jurg}(2009)}]{Marx:2009}%
  \BibitemOpen
  \bibfield  {author} {\bibinfo {author} {\bibfnamefont {D.}~\bibnamefont
  {Marx}}\ and\ \bibinfo {author} {\bibfnamefont {H.}~\bibnamefont {Jurg}},\
  }\href@noop {} {\emph {\bibinfo {title} {Ab Initio Molecular Dynamics: Basic
  Theory and Advanced Methods}}}\ (\bibinfo  {publisher} {Cambridge University
  Press},\ \bibinfo {year} {2009})\BibitemShut {NoStop}%
\bibitem [{\citenamefont {Ehrenfest}(1927)}]{Ehrenfest:1927}%
  \BibitemOpen
  \bibfield  {author} {\bibinfo {author} {\bibfnamefont {P.}~\bibnamefont
  {Ehrenfest}},\ }\href {\doibase 10.1007/BF01329203} {\bibfield  {journal}
  {\bibinfo  {journal} {Z. Phys. A Hadron. Nucl.}\ }\textbf {\bibinfo {volume}
  {45}},\ \bibinfo {pages} {455} (\bibinfo {year} {1927})}\BibitemShut
  {NoStop}%
\bibitem [{\citenamefont {Schleife}\ \emph {et~al.}(2014)\citenamefont
  {Schleife}, \citenamefont {Draeger}, \citenamefont {Anisimov}, \citenamefont
  {Correa},\ and\ \citenamefont {Kanai}}]{Schleife:2014}%
  \BibitemOpen
  \bibfield  {author} {\bibinfo {author} {\bibfnamefont {A.}~\bibnamefont
  {Schleife}}, \bibinfo {author} {\bibfnamefont {E.~W.}\ \bibnamefont
  {Draeger}}, \bibinfo {author} {\bibfnamefont {V.~M.}\ \bibnamefont
  {Anisimov}}, \bibinfo {author} {\bibfnamefont {A.~A.}\ \bibnamefont
  {Correa}}, \ and\ \bibinfo {author} {\bibfnamefont {Y.}~\bibnamefont
  {Kanai}},\ }\href {\doibase 10.1109/MCSE.2014.55} {\bibfield  {journal}
  {\bibinfo  {journal} {Comput. Sci. Eng.}\ }\textbf {\bibinfo {volume} {16}},\
  \bibinfo {pages} {54} (\bibinfo {year} {2014})}\BibitemShut {NoStop}%
\bibitem [{\citenamefont {Draeger}\ \emph {et~al.}(2017)\citenamefont
  {Draeger}, \citenamefont {Andrade}, \citenamefont {Gunnels}, \citenamefont
  {Bhatele}, \citenamefont {Schleife},\ and\ \citenamefont
  {Correa}}]{Draeger:2017}%
  \BibitemOpen
  \bibfield  {author} {\bibinfo {author} {\bibfnamefont {E.~W.}\ \bibnamefont
  {Draeger}}, \bibinfo {author} {\bibfnamefont {X.}~\bibnamefont {Andrade}},
  \bibinfo {author} {\bibfnamefont {J.~A.}\ \bibnamefont {Gunnels}}, \bibinfo
  {author} {\bibfnamefont {A.}~\bibnamefont {Bhatele}}, \bibinfo {author}
  {\bibfnamefont {A.}~\bibnamefont {Schleife}}, \ and\ \bibinfo {author}
  {\bibfnamefont {A.~A.}\ \bibnamefont {Correa}},\ }\href {\doibase
  10.1016/j.jpdc.2017.02.005} {\bibfield  {journal} {\bibinfo  {journal} {J.
  Parallel Distr. Com.}\ }\textbf {\bibinfo {volume} {106}},\ \bibinfo {pages}
  {205} (\bibinfo {year} {2017})}\BibitemShut {NoStop}%
\bibitem [{\citenamefont {Runge}\ and\ \citenamefont
  {Gross}(1984)}]{Runge1984}%
  \BibitemOpen
  \bibfield  {author} {\bibinfo {author} {\bibfnamefont {E.}~\bibnamefont
  {Runge}}\ and\ \bibinfo {author} {\bibfnamefont {E.~K.~U.}\ \bibnamefont
  {Gross}},\ }\href {\doibase 10.1103/PhysRevLett.52.997} {\bibfield  {journal}
  {\bibinfo  {journal} {Phys. Rev. Lett.}\ }\textbf {\bibinfo {volume} {52}},\
  \bibinfo {pages} {997} (\bibinfo {year} {1984})}\BibitemShut {NoStop}%
\bibitem [{\citenamefont {Schleife}\ \emph {et~al.}(2012)\citenamefont
  {Schleife}, \citenamefont {Draeger}, \citenamefont {Kanai},\ and\
  \citenamefont {Correa}}]{Schleife:2012_c}%
  \BibitemOpen
  \bibfield  {author} {\bibinfo {author} {\bibfnamefont {A.}~\bibnamefont
  {Schleife}}, \bibinfo {author} {\bibfnamefont {E.~W.}\ \bibnamefont
  {Draeger}}, \bibinfo {author} {\bibfnamefont {Y.}~\bibnamefont {Kanai}}, \
  and\ \bibinfo {author} {\bibfnamefont {A.~A.}\ \bibnamefont {Correa}},\
  }\href {\doibase 10.1063/1.4758792} {\bibfield  {journal} {\bibinfo
  {journal} {J. Chem. Phys.}\ }\textbf {\bibinfo {volume} {137}},\ \bibinfo
  {eid} {22A546} (\bibinfo {year} {2012})}\BibitemShut {NoStop}%
\bibitem [{\citenamefont {Ziegler}(1980)}]{Ziegler:1980}%
  \BibitemOpen
  \bibfield  {author} {\bibinfo {author} {\bibfnamefont {J.~F.}\ \bibnamefont
  {Ziegler}},\ }\href@noop {} {\emph {\bibinfo {title} {Handbook of stopping
  cross-sections for energetic ions in all elements}}}\ (\bibinfo  {publisher}
  {Pergamon Press},\ \bibinfo {address} {New York},\ \bibinfo {year} {1980})\
  p.\ \bibinfo {pages} {432}\BibitemShut {NoStop}%
\bibitem [{\citenamefont {Ziegler}\ \emph {et~al.}(2010)\citenamefont
  {Ziegler}, \citenamefont {Ziegler},\ and\ \citenamefont
  {Biersack}}]{Ziegler2010}%
  \BibitemOpen
  \bibfield  {author} {\bibinfo {author} {\bibfnamefont {J.~F.}\ \bibnamefont
  {Ziegler}}, \bibinfo {author} {\bibfnamefont {M.~D.}\ \bibnamefont
  {Ziegler}}, \ and\ \bibinfo {author} {\bibfnamefont {J.~P.}\ \bibnamefont
  {Biersack}},\ }\href {\doibase 10.1016/j.nimb.2010.02.091} {\bibfield
  {journal} {\bibinfo  {journal} {Nucl. Instrum. Meth. B}\ }\textbf {\bibinfo
  {volume} {268}},\ \bibinfo {pages} {1818} (\bibinfo {year}
  {2010})}\BibitemShut {NoStop}%
\bibitem [{\citenamefont {Smith}\ and\ \citenamefont {Webb}(1991)}]{Smith1991}%
  \BibitemOpen
  \bibfield  {author} {\bibinfo {author} {\bibfnamefont {R.}~\bibnamefont
  {Smith}}\ and\ \bibinfo {author} {\bibfnamefont {R.~P.}\ \bibnamefont
  {Webb}},\ }\href {\doibase 10.1080/09500839108214619} {\bibfield  {journal}
  {\bibinfo  {journal} {Phil. Mag. Lett.}\ }\textbf {\bibinfo {volume} {64}},\
  \bibinfo {pages} {253} (\bibinfo {year} {1991})}\BibitemShut {NoStop}%
\bibitem [{\citenamefont {Schleife}\ \emph {et~al.}(2015)\citenamefont
  {Schleife}, \citenamefont {Kanai},\ and\ \citenamefont
  {Correa}}]{Schleife2015}%
  \BibitemOpen
  \bibfield  {author} {\bibinfo {author} {\bibfnamefont {A.}~\bibnamefont
  {Schleife}}, \bibinfo {author} {\bibfnamefont {Y.}~\bibnamefont {Kanai}}, \
  and\ \bibinfo {author} {\bibfnamefont {A.~A.}\ \bibnamefont {Correa}},\
  }\href {\doibase 10.1103/PhysRevB.91.014306} {\bibfield  {journal} {\bibinfo
  {journal} {Phys. Rev. B}\ }\textbf {\bibinfo {volume} {91}},\ \bibinfo
  {pages} {014306} (\bibinfo {year} {2015})}\BibitemShut {NoStop}%
\bibitem [{\citenamefont {Ullah}\ \emph {et~al.}(2015)\citenamefont {Ullah},
  \citenamefont {Corsetti}, \citenamefont {S\'anchez-Portal},\ and\
  \citenamefont {Artacho}}]{Ullah:2015}%
  \BibitemOpen
  \bibfield  {author} {\bibinfo {author} {\bibfnamefont {R.}~\bibnamefont
  {Ullah}}, \bibinfo {author} {\bibfnamefont {F.}~\bibnamefont {Corsetti}},
  \bibinfo {author} {\bibfnamefont {D.}~\bibnamefont {S\'anchez-Portal}}, \
  and\ \bibinfo {author} {\bibfnamefont {E.}~\bibnamefont {Artacho}},\ }\href
  {\doibase 10.1103/PhysRevB.91.125203} {\bibfield  {journal} {\bibinfo
  {journal} {Phys. Rev. B}\ }\textbf {\bibinfo {volume} {91}},\ \bibinfo
  {pages} {125203} (\bibinfo {year} {2015})}\BibitemShut {NoStop}%
\bibitem [{\citenamefont {Winter}\ \emph {et~al.}(2003)\citenamefont {Winter},
  \citenamefont {Juaristi}, \citenamefont {Nagy}, \citenamefont {Arnau},\ and\
  \citenamefont {Echenique}}]{Winter:2003}%
  \BibitemOpen
  \bibfield  {author} {\bibinfo {author} {\bibfnamefont {H.}~\bibnamefont
  {Winter}}, \bibinfo {author} {\bibfnamefont {J.~I.}\ \bibnamefont
  {Juaristi}}, \bibinfo {author} {\bibfnamefont {I.}~\bibnamefont {Nagy}},
  \bibinfo {author} {\bibfnamefont {A.}~\bibnamefont {Arnau}}, \ and\ \bibinfo
  {author} {\bibfnamefont {P.~M.}\ \bibnamefont {Echenique}},\ }\href {\doibase
  10.1103/PhysRevB.67.245401} {\bibfield  {journal} {\bibinfo  {journal} {Phys.
  Rev. B}\ }\textbf {\bibinfo {volume} {67}},\ \bibinfo {pages} {245401}
  (\bibinfo {year} {2003})}\BibitemShut {NoStop}%
\bibitem [{\citenamefont {Quashie}\ \emph {et~al.}(2016)\citenamefont
  {Quashie}, \citenamefont {Saha},\ and\ \citenamefont
  {Correa}}]{Quashie:2016}%
  \BibitemOpen
  \bibfield  {author} {\bibinfo {author} {\bibfnamefont {E.~E.}\ \bibnamefont
  {Quashie}}, \bibinfo {author} {\bibfnamefont {B.~C.}\ \bibnamefont {Saha}}, \
  and\ \bibinfo {author} {\bibfnamefont {A.~A.}\ \bibnamefont {Correa}},\
  }\href {\doibase 10.1103/PhysRevB.94.155403} {\bibfield  {journal} {\bibinfo
  {journal} {Phys. Rev. B}\ }\textbf {\bibinfo {volume} {94}},\ \bibinfo
  {pages} {155403} (\bibinfo {year} {2016})}\BibitemShut {NoStop}%
\bibitem [{\citenamefont {Yost}\ \emph {et~al.}(2017)\citenamefont {Yost},
  \citenamefont {Yao},\ and\ \citenamefont {Kanai}}]{Yost:2017}%
  \BibitemOpen
  \bibfield  {author} {\bibinfo {author} {\bibfnamefont {D.~C.}\ \bibnamefont
  {Yost}}, \bibinfo {author} {\bibfnamefont {Y.}~\bibnamefont {Yao}}, \ and\
  \bibinfo {author} {\bibfnamefont {Y.}~\bibnamefont {Kanai}},\ }\href
  {\doibase 10.1103/PhysRevB.96.115134} {\bibfield  {journal} {\bibinfo
  {journal} {Phys. Rev. B}\ }\textbf {\bibinfo {volume} {96}},\ \bibinfo
  {pages} {115134} (\bibinfo {year} {2017})}\BibitemShut {NoStop}%
\bibitem [{\citenamefont {Nazarov}\ \emph {et~al.}(2007)\citenamefont
  {Nazarov}, \citenamefont {Pitarke}, \citenamefont {Takada}, \citenamefont
  {Vignale},\ and\ \citenamefont {Chang}}]{Nazarov:2007}%
  \BibitemOpen
  \bibfield  {author} {\bibinfo {author} {\bibfnamefont {V.~U.}\ \bibnamefont
  {Nazarov}}, \bibinfo {author} {\bibfnamefont {J.~M.}\ \bibnamefont
  {Pitarke}}, \bibinfo {author} {\bibfnamefont {Y.}~\bibnamefont {Takada}},
  \bibinfo {author} {\bibfnamefont {G.}~\bibnamefont {Vignale}}, \ and\
  \bibinfo {author} {\bibfnamefont {Y.-C.}\ \bibnamefont {Chang}},\ }\href
  {\doibase 10.1103/PhysRevB.76.205103} {\bibfield  {journal} {\bibinfo
  {journal} {Phys. Rev. B}\ }\textbf {\bibinfo {volume} {76}},\ \bibinfo
  {pages} {205103} (\bibinfo {year} {2007})}\BibitemShut {NoStop}%
\bibitem [{\citenamefont {Correa}(2018)}]{Correa:2018}%
  \BibitemOpen
  \bibfield  {author} {\bibinfo {author} {\bibfnamefont {A.~A.}\ \bibnamefont
  {Correa}},\ }\href {\doibase https://doi.org/10.1016/j.commatsci.2018.03.064}
  {\bibfield  {journal} {\bibinfo  {journal} {Comput. Mater. Sci.}\ }\textbf
  {\bibinfo {volume} {150}},\ \bibinfo {pages} {291 } (\bibinfo {year}
  {2018})}\BibitemShut {NoStop}%
\bibitem [{\citenamefont {Lindhard~J.}(1964)}]{Lindhard:1964}%
  \BibitemOpen
  \bibfield  {author} {\bibinfo {author} {\bibfnamefont {W.~A.}\ \bibnamefont
  {Lindhard~J.}},\ }\href@noop {} {\bibfield  {journal} {\bibinfo  {journal}
  {Mat. Fys. Medd. Dan. Vid. Selsk.}\ }\textbf {\bibinfo {volume} {34}},\
  \bibinfo {pages} {1} (\bibinfo {year} {1964})}\BibitemShut {NoStop}%
\bibitem [{\citenamefont {Seddiki}\ \emph {et~al.}(2013)\citenamefont
  {Seddiki}, \citenamefont {Ouahrani}, \citenamefont {Lasri}, \citenamefont
  {Benouaz}, \citenamefont {Reshak},\ and\ \citenamefont
  {Bouhafs}}]{SEDDIKI2013}%
  \BibitemOpen
  \bibfield  {author} {\bibinfo {author} {\bibfnamefont {N.}~\bibnamefont
  {Seddiki}}, \bibinfo {author} {\bibfnamefont {T.}~\bibnamefont {Ouahrani}},
  \bibinfo {author} {\bibfnamefont {B.}~\bibnamefont {Lasri}}, \bibinfo
  {author} {\bibfnamefont {T.}~\bibnamefont {Benouaz}}, \bibinfo {author}
  {\bibfnamefont {A.}~\bibnamefont {Reshak}}, \ and\ \bibinfo {author}
  {\bibfnamefont {B.}~\bibnamefont {Bouhafs}},\ }\href {\doibase
  https://doi.org/10.1016/j.mssp.2013.04.006} {\bibfield  {journal} {\bibinfo
  {journal} {Mat. Sci. in Semicon. Proc.}\ }\textbf {\bibinfo {volume} {16}},\
  \bibinfo {pages} {1454 } (\bibinfo {year} {2013})}\BibitemShut {NoStop}%
\bibitem [{\citenamefont {Correa}\ \emph {et~al.}(2012)\citenamefont {Correa},
  \citenamefont {Kohanoff}, \citenamefont {Artacho}, \citenamefont
  {S\'anchez-Portal},\ and\ \citenamefont {Caro}}]{Correa:2012}%
  \BibitemOpen
  \bibfield  {author} {\bibinfo {author} {\bibfnamefont {A.~A.}\ \bibnamefont
  {Correa}}, \bibinfo {author} {\bibfnamefont {J.}~\bibnamefont {Kohanoff}},
  \bibinfo {author} {\bibfnamefont {E.}~\bibnamefont {Artacho}}, \bibinfo
  {author} {\bibfnamefont {D.}~\bibnamefont {S\'anchez-Portal}}, \ and\
  \bibinfo {author} {\bibfnamefont {A.}~\bibnamefont {Caro}},\ }\href {\doibase
  10.1103/PhysRevLett.108.213201} {\bibfield  {journal} {\bibinfo  {journal}
  {Phys. Rev. Lett.}\ }\textbf {\bibinfo {volume} {108}},\ \bibinfo {pages}
  {213201} (\bibinfo {year} {2012})}\BibitemShut {NoStop}%
\bibitem [{\citenamefont {Gemmell}(1974)}]{Gemmell:1974}%
  \BibitemOpen
  \bibfield  {author} {\bibinfo {author} {\bibfnamefont {D.~S.}\ \bibnamefont
  {Gemmell}},\ }\href {\doibase 10.1103/RevModPhys.46.129} {\bibfield
  {journal} {\bibinfo  {journal} {Rev. Mod. Phys.}\ }\textbf {\bibinfo {volume}
  {46}},\ \bibinfo {pages} {129} (\bibinfo {year} {1974})}\BibitemShut
  {NoStop}%
\bibitem [{\citenamefont {Lim}\ \emph {et~al.}(2016)\citenamefont {Lim},
  \citenamefont {Foulkes}, \citenamefont {Horsfield}, \citenamefont {Mason},
  \citenamefont {Schleife}, \citenamefont {Draeger},\ and\ \citenamefont
  {Correa}}]{Lim:2016}%
  \BibitemOpen
  \bibfield  {author} {\bibinfo {author} {\bibfnamefont {A.}~\bibnamefont
  {Lim}}, \bibinfo {author} {\bibfnamefont {W.~M.~C.}\ \bibnamefont {Foulkes}},
  \bibinfo {author} {\bibfnamefont {A.~P.}\ \bibnamefont {Horsfield}}, \bibinfo
  {author} {\bibfnamefont {D.~R.}\ \bibnamefont {Mason}}, \bibinfo {author}
  {\bibfnamefont {A.}~\bibnamefont {Schleife}}, \bibinfo {author}
  {\bibfnamefont {E.~W.}\ \bibnamefont {Draeger}}, \ and\ \bibinfo {author}
  {\bibfnamefont {A.~A.}\ \bibnamefont {Correa}},\ }\href {\doibase
  10.1103/PhysRevLett.116.043201} {\bibfield  {journal} {\bibinfo  {journal}
  {Phys. Rev. Lett.}\ }\textbf {\bibinfo {volume} {116}},\ \bibinfo {pages}
  {043201} (\bibinfo {year} {2016})}\BibitemShut {NoStop}%
\bibitem [{\citenamefont {Kramida}\ \emph {et~al.}(2018)\citenamefont
  {Kramida}, \citenamefont {Ralchenko}, \citenamefont {Reader},\ and\
  \citenamefont {{NIST ASD Team}}}]{NIST2018}%
  \BibitemOpen
  \bibfield  {author} {\bibinfo {author} {\bibfnamefont {A.}~\bibnamefont
  {Kramida}}, \bibinfo {author} {\bibfnamefont {Y.}~\bibnamefont {Ralchenko}},
  \bibinfo {author} {\bibfnamefont {J.}~\bibnamefont {Reader}}, \ and\ \bibinfo
  {author} {\bibnamefont {{NIST ASD Team}}},\ }\href
  {https://physics.nist.gov/asd} {\enquote {\bibinfo {title} {Nist atomic
  spectra database (version 5.5.3)},}\ } (\bibinfo {year} {2018}),\ \bibinfo
  {note} {{National Institute of Standards and Technology}}\BibitemShut
  {NoStop}%
\end{thebibliography}%

\newpage
\onecolumngrid

\section{Supplemental Material}

\subsection{Estimation of threshold velocity for excitation due to fast charged particle}

As discussed in Ref.\ \onlinecite{Lim:2016}, the threshold velocity, below which no electronic stopping is allowed, can be estimated by Planck's constant ($h$), distance between equivalent lattice position ($\lambda$), and band gap ($\Delta$),
\begin{equation}
\label{eq:vth}
v_{th}=\frac{\lambda\Delta}{h}.
\end{equation}
We extend Eq.\ \eqref{eq:vth} to estimate the threshold velocity to excite electrons from each shell to the conduction band minimum by replacing the band gap with the corresponding energy difference, calculated by subtracting electron affinity from ionization energy. 
The distance between equivalent lattice positions is 1/2 of a lattice period, i.e., 1.55, 1.43, and 1.50 $a_\mathrm{B}$ for InP, GaP, and In$_{0.5}$Ga$_{0.5}$P, respectively.
Since this is only an estimate, 1.50 $a_\mathrm{B}$ is used for all the calculations.
The electron affinity for InP, GaP, and In$_{0.5}$Ga$_{0.5}$P is 0.16, 0.14, and 0.15 $E_\mathrm{H}$, respectively, and 0.15 $E_\mathrm{H}$ is used for all the calculations.
Ionization energy and threshold velocity (kinetic energy) for each shell are shown in Table \ref{vth_all}.
Note that since the estimation is based on atomic spectral data and intra-band excitations within valence electrons are not considered, it can only serve as rough estimation and all-electron calculation is ultimately needed to study the contribution of semi-core electrons.

\begin{table}[h!]
\caption{\label{vth_all}Threshold velocity of each shell based on ionization energy\cite{NIST2018} for In, Ga, and P atom. 
The first semi-core levels that are not included in pseudopotentials are marked in bold.
}
\begin{tabular}{ | l | l | l | l || l | l | l | l || l | l | l | l | }
\hline
	In &  &  &  & Ga &  &  &  & P &  &  &  \\ 
	shell & I.E.\ (eV) & $v_\mathrm{th}$ (at.\ u.) & K.E.$_\mathrm{th}$ (keV) & shell & I.E.\ (eV) & $v_\mathrm{th}$ (at.\ u.) & K.E.$_\mathrm{th}$ (keV) & shell & I.E.\ (eV) & $v_\mathrm{th}$ (at.\ u.) & K.E.$_\mathrm{th}$ (keV) \\ \hline
	5\,$p$ & 5.78 & 0.053 & 0.071 & 4\,$p$ & 5.99 & 0.059 & 0.088 & 3\,$p$ & 10.49 & 0.20 & 1.00 \\
	5\,$s$ & 18.87 & 0.47 & 5.52 & 4\,$s$ & 20.52& 0.52 & 6.76 &  & 19.77 & 0.50 & 6.25 \\ 
	 & 28.04 & 0.75 & 14.1 &  & 30.72 & 0.84 & 17.6 &  & 30.2 & 0.82 & 16.8 \\ 
	4\,$d$ & 55.45 & 1.61 & 64.8 & 3\,$d$ & 63.241 & 1.85 & 85.6 & 3\,$s$ & 51.44 & 1.49 & 55.5 \\ 
	 & 69.31 & 2.04 & 105 &  & 86.01 & 2.57 & 166 &  & 65.03 & 1.91 & 92 \\ 
	 & 90 & 2.69 & 181 &  & 112.7 & 3.40 & 289 & 2\,$p$ & \textbf{220.43} & \textbf{6.77} & \textbf{1146} \\ 
	 & 109 & 3.29 & 271 &  & 140.8 & 4.28 & 458 &  & 263.57 & 8.12 & 1649 \\ 
	 & 130.1 & 3.95 & 391 &  & 169.9 & 5.19 & 674 &  & 309.60 & 9.56 & 2285 \\
	 & 156 & 4.76 & 567 &  & 211 & 6.48 & 1050 &  & 372.31 & 11.52 & 3318 \\ 
	 & 178 & 5.44 & 740 &  & 244 & 7.51 & 1410 &  & 424.4 & 13.15 & 4323 \\ 
	 & 201 & 6.16 & 949 &  & 280 & 8.63 & 1862 &  & 479.44 & 14.87 & 5528 \\ 
	 & 226 & 6.95 & 1208 &  & 319 & 9.85 & 2426 & 2\,$s$ & 560.62 & 17.41 & 7578 \\ 
	 & 249 & 7.67 & 1471 &  & 356 & 11.01 & 3031 &  & 611.74 & 19.01 & 9034 \\ 
	4\,$p$ & \textbf{341} & \textbf{10.54} & \textbf{2778} & 3\,$p$ & \textbf{471.2} & \textbf{14.62} & \textbf{5344} &  &  &  &  \\ 
	 & 368 & 11.39 & 3244 &  & 508.6 & 15.79 & 6233 &  &  &  &  \\ 
	 & 396 & 12.26 & 3758 &  & 548.3& 17.03 & 7251 &  &  &  &  \\ 
	 & 425 & 13.17 & 4336 &  & 599.8 & 18.64 & 8686 &  &  &  &  \\ 
	 & 462 & 14.33 & 5134 &  & 640 & 19.90 & 9900 &  &  &  &  \\ 
	 & 497 & 15.42 & 5944 &  & 676.9 & 21.05 & 11077 &  &  &  &  \\ 
	4\,$s$ & 560 & 17.39 & 7560 & 3\,$s$ & 765.7 & 23.83 & 14196 &  &  &  &  \\ 
	 & 593.3 & 18.43 & 8492 &  & 807.3 & 25.13 & 15787 &  &  &  &  \\ \hline
\end{tabular}
\end{table}

\subsection{\label{sec:off_ch_dis}Calculation of electronic stopping for an off-channeling trajectory}

\begin{figure}[ht!]
\includegraphics[width=0.95\columnwidth]{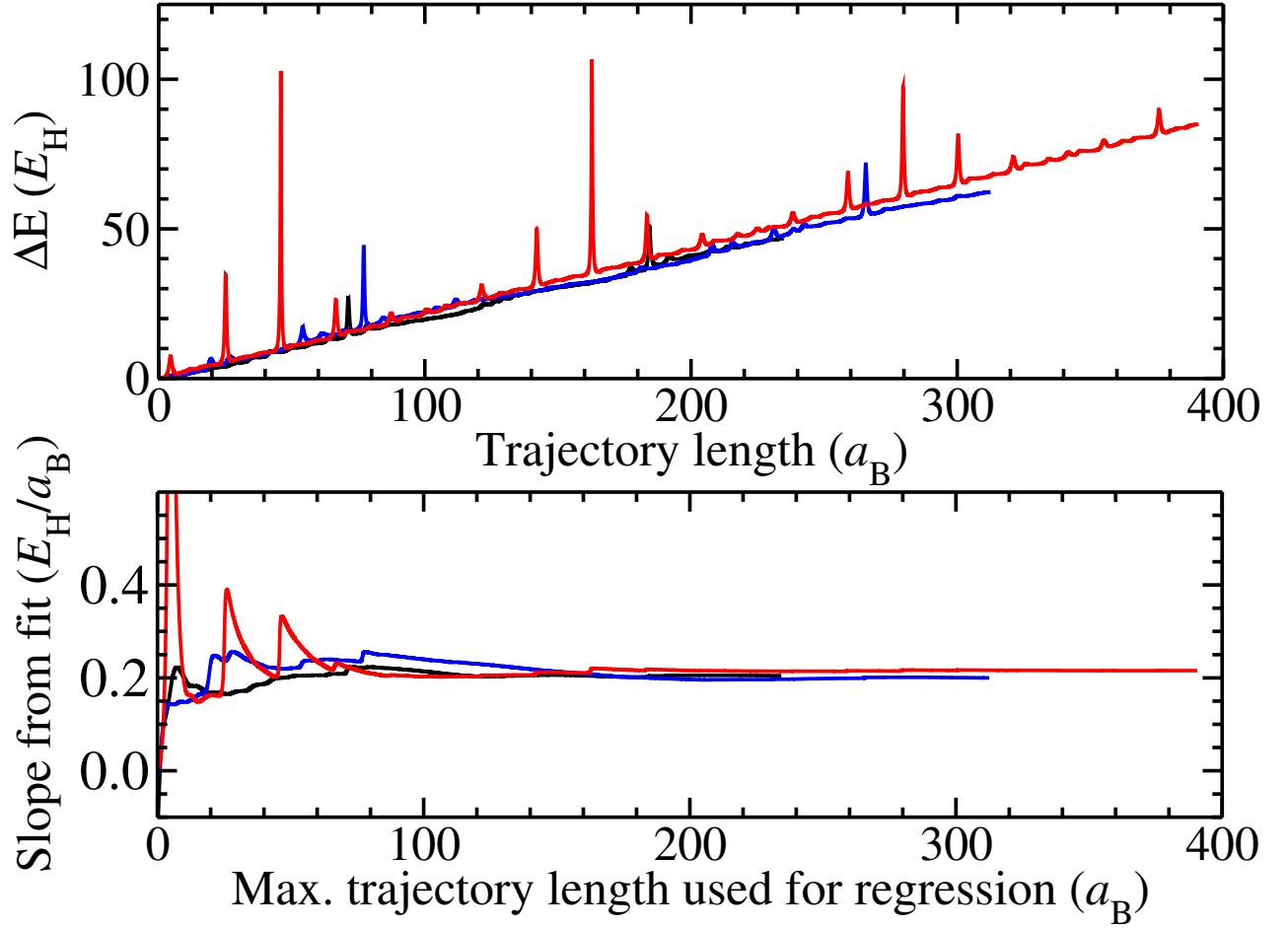}
\caption{\label{fig:off_channel}Convergence of electronic stopping power of GaP (black), $\mathrm{In_{0.5}Ga_{0.5}P}$ (blue), and InP (red) for a proton at velocity of 0.9 at.\ u.\ with off-channeling trajectory. Top subfigure is the energy gain along the trajectory while bottom sub-figure is the regression fit of given maximum trajectory length.
}
\end{figure}

The energy transfer from proton to the target material depends on the local environment and is trajectory dependent.
When a proton is closer to nuclei of the target material, it has higher probability to excite electrons since the electron density is higher. 
The proton also has higher chance to excite core electrons for the same reason.
Therefore, the energy transferred from proton to target materials is larger when proton travels near positions of nuclei.
While the shortest distance between proton and target ion leads to the sharpest peaks in Fig.\ \ref{fig:off_channel}, we also note that these results are affected by the cutoff radius of the pseudopotentials used here.
For this reason, we only use the average to extract stopping, as explained in detail in Ref.\ \onlinecite{Schleife2015}.

Nevertheless, counting the peaks in each trajectory, we clearly find that there are much more and higher peaks for the trajectory, on which proton travels in InP, than the other two trajectories.
This indicates that for the short trajectory we used to calculate the electronic stopping of InP, the proton happens to experience region of higher electron density.
Therefore, we predict higher electronic stopping than fully converged value for InP.
Decreasing height of the peaks for InP also suggest that the trajectory start to explore region of lower electron density.
Therefore, we expect a much longer trajectory can have better sampling of the target materials and thus predict electronic stopping closer to converged value.
 
\subsection{Error estimate for channeling projectiles}

\begin{table}[h!]
\caption{\label{tab:SP_error}Numerical error due to choice of region to average (in lattice periods). Error is calculated using 0.5\,--\,2.5 as reference, since in the manuscript we discard the first and last half period (see main text). This data is for GaP.
}
\begin{tabular}{| l | l | l | l | l || l | l | l | l | l |}
\hline
	$v$ [100] & 0.5\,--\,2.5 & 0.5\,--\,1.5 & 1.5\,--\,2.5 & 1.0\,--\,2.0 & $v$ [110] & 0.5\,--\,2.5 & 0.5\,--\,1.5 & 1.5\,--\,2.5 & 1.0\,--\,2.0 \\ \hline
	0.2 & 3.65E-2 & 3.66E-2 & 3.64E-2 & 3.67E-2 & 0.2 & 3.26E-2 & 3.29E-2 & 3.23E-2 & 3.25E-2 \\ \hline
	0.5 & 0.112 & 0.113 & 0.1114 & 0.112 & 0.5 & 7.99E-2 & 8.04E-2 & 7.96E-2 & 8.02E-2 \\ \hline
	1.5 & 0.255 & 0.268 & 0.243 & 0.249 & 1.5 & 0.137 & 0.143 & 0.132 & 0.134 \\ \hline
	2.5 & 0.171 & 0.169 & 0.173 & 0.172 & 2.5 & 8.94E-2 & 8.97E-2 & 8.92E-2 & 8.98E-2 \\ \hline
	3 & 0.133 & 0.130 & 0.136 & 0.133 & 3 & 7.00E-2 & 6.95E-2 & 7.04E-2 & 7.05E-2 \\ \hline
	4 & 8.39E-2 & 8.22E-2 & 8.57E-2 & 8.48E-2 & 4 & 4.62E-2 & 4.50E-2 & 4.75E-2 & 4.64-2 \\ \hline
	5 & 5.72E-2 & 5.67E-2 & 5.77E-2 & 5.87E-2 & 5 & 3.27E-2 & 3.17E-2 & 3.37E-2 & 3.30E-2 \\ \hline
	\ & error (\%) & \  & \  & \  & \ & error (\%)  & \  & \  & \  \\ \hline
	0.2 & \ & 0.233 & $-0.233$ & 0.473 & 0.2 & \ & 0.904 & $-0.904$ & $-0.239$   \\ \hline
	0.5 & \ & 0.843 & $-0.843$ & $-0.141$ & 0.5 & \ & 0.507 & $-0.507$ & 0.283   \\ \hline
	1.5 & \ &  4.90 & $-4.90$ & $-2.49$ & 1.5 & \ & 3.89 & $-3.89$ & $-2.75$  \\ \hline
	2.5 & \ &  $-1.23$ & 1.24 & 0.56 & 2.5 & \ & 0.31 & $-0.31$ & 0.46  \\ \hline
	3 & \ & $-2.15$ & 2.15 & 0.441 & 3 & \ & $-0.65$ & 0.65 & 0.71   \\ \hline
	4 & \ & $-2.12$ & 2.12 & 1.03 & 4 & \ & $-2.64$ & 2.64 & 0.39   \\ \hline
	5 & \ & $-0.941$ & 0.94 & 2.57 & 5 & \ & $-3.10$ & 3.10 & 0.86    \\ \hline
\end{tabular}
\end{table}

\end{document}